\documentclass[letterpaper, 11pt]{scrartcl}
\usepackage[latin1]{inputenc}
\usepackage[T1]{fontenc}
\usepackage{graphicx}
\usepackage{graphics}
\usepackage{verbatim}
\usepackage{amsmath}
\usepackage{amssymb}
\usepackage{amsfonts}
\usepackage{dsfont}
\usepackage{fancybox}
\usepackage{booktabs}
\usepackage[left=2.5cm,right=2.5cm,top=2.5cm,bottom=2.5cm]{geometry}
\usepackage{hhline}
\usepackage{longtable}
\usepackage{multirow}
\usepackage{rotating}
\numberwithin{equation}{section}
\numberwithin{table}{section}

\newcommand{\F}{\mathcal{F}}
\newcommand{\cL}{\mathcal{L}}
\newcommand{\M}{\mathcal{M}}
\newcommand{\cO}{\mathcal{O}}

\newcommand{\cF}{{\cal F}}
\newcommand{\cD}{{\cal D}}
\newcommand{\cM}{{\cal M}}

\newcommand{\cZ}{{\cal Z}}

\def\itau{{\rm Im}(\tau)}
\renewcommand{\d}{\partial}
\newcommand{\dbar}{\overline{\partial}}

\newcommand{\cref}{{\bf [check ref]}}

\newcommand{\half}{\frac12}
\newcommand{\quart}{\frac14}
\newcommand{\dd}{{\rm d}}
\renewcommand{\S}{\Sigma}
\newcommand{\HirzeF}{\mathds{F}}

\newcommand{\be}{\begin{equation}}
\newcommand{\ee}{\end{equation}}
\newcommand{\bea}{\begin{eqnarray}}
\newcommand{\eea}{\end{eqnarray}}

\newcommand{\Z}{\mathds{Z}}

\newcommand{\R}{\mathds{R}}
\newcommand{\C}{\mathds{C}}
\renewcommand{\P}{\mathds{P}}
\newcommand{\K}{\mathds{K}}
\newcommand{\B}{\mathds{B}}
\renewcommand{\H}{\mathds{H}}

\newcommand{\bi}{{\bar\imath}}

\newcommand{\cz}[1]{\ensuremath{z_{c,#1}}} 
\newcommand{\ct}[1]{\ensuremath{t_{c,#1}}} 
\newcommand{\oz}[1]{\ensuremath{z_{o,#1}}} 
\newcommand{\ot}[1]{\ensuremath{t_{o,#1}}} 
\newcommand{\ott}[1]{\ensuremath{\tilde{t}_{o,#1}}} 
\newcommand{\xT}[1]{\ensuremath{t_{x,#1}}} 

\begin{document}


\baselineskip=18pt

\begin{titlepage}
\begin{flushright}
\parbox[t]{1.8in}{
BONN-TH-2008-10
}
\end{flushright}

\begin{center}

\vspace*{ 1.2cm}

{\large \bf \Large Integrability of the holomorphic anomaly equations}

\vskip 1.2cm

\begin{center}
 \bf{Babak Haghighat\footnote{~babak@th.physik.uni-bonn.de}, Albrecht Klemm\footnote{~aklemm@th.physik.uni-bonn.de} and Marco Rauch\footnote{~rauch@th.physik.uni-bonn.de}}
\end{center}
\vskip 0.2cm

{\em Physikalisches Institut der Universit\"at Bonn \\[.1cm]
and\\[.1cm]
Bethe Center for Theoretical Physics \\[.1cm]
Nu\ss allee 12, 53115 Bonn, Germany}
\vspace*{.1cm}

\end{center}

\vskip 0.2cm

\begin{center} {\bf ABSTRACT } \end{center}
We show that modularity and the gap condition make the holomorphic anomaly equation 
completely integrable for non-compact Calabi-Yau manifolds. This leads to a very efficient
formalism to solve the topological string on these geometries in terms of almost 
holomorphic modular forms. The formalism provides in particular holomorphic expansions 
everywhere in moduli space including large radius points, the conifold loci, 
Seiberg-Witten points and the orbifold points. It can be also viewed as a very 
efficient method to solve higher genus closed string amplitudes in the
$\frac{1}{N^2}$ expansion of  matrix models with more then one cut.

\end{titlepage}


\tableofcontents
\newpage
\section{Introduction}

String theory on non-compact Calabi-Yau geometries is relevant for the construction of 
4d supersymmetric theories decoupled from gravity and provides simple examples for important 
concepts of string theory in nontrivial geometrical backgrounds, as e.g. the behavior of 
the amplitudes under topology change of the background geometry.  
Exploring the topological sector has been especially fruitful in providing examples of large 
$N$-dualities connecting topological string theory on these backgrounds to 3d Chern-Simons 
theory and matrix models. If the geometric background has a non-trivial space time duality symmetry 
group, which is the case if the local mirror geometry involves a Riemann surface
of at least genus one, the situation is as follows. 
Large $N$-dualities or localization principles apply  to certain  holomorphic limits of 
the topological string amplitudes and lead to local holomorphic expansion of 
the latter at special points in the moduli space of the theory. 
Typically at large radius these come in closed formulas involving infinite sums or
products over partitions coming from joining topological vertices or from
Nekrasov localization formulas. The expressions lead to formal,
i.e. non-convergent expansions, in the string coupling whose coefficients 
have finite radius of convergence in the moduli parameter. 
However, since these limits break the invariance of the amplitudes under the space 
duality group this fundamental symmetry property of the theory is obscured.  

In this article we show that a simple bootstrap approach using extensively 
the full space time modular  invariance, the holomorphic anomaly equation and
a local analysis of the gap condition at the nodes is highly efficient in 
reconstructing modular invariant, non-holomorphic string  amplitudes for local 
Calabi-Yau spaces to all genus. They are polynomials in 
generators of the modular groups, which are globally defined in the moduli 
space of the theory. As a consequence the amplitudes are globally defined and 
holomorphic limits can be easily obtained everywhere in the moduli space. The
approach extends to $N=2$ gauge theories and matrix models.    

The paper is organized as follows. In section \ref{localmirrorsymmetry} we 
recall the local Calabi-Yau A-model geometries and how local mirror symmetry 
leads to a B-model geometry that is governed by 
a family of Riemann surfaces $\S_g$ with a canonical meromorphic differential. 
We derive the  Picard-Fuchs equations for the periods and their solutions and thereby 
solve the genus zero sector.          
   
In section \ref{directintegration} we discuss the formalism of direct 
integration for local Calabi-Yau spaces. The space-time modular group of $\S_g$ 
is a finite index  subgroup $\Gamma$ of ${\rm Sp}(2g,\Z)$. The invariance of 
the closed topological string amplitudes $F_g$ under $\Gamma$ and the holomorphic anomaly 
equation implies that the $F_g$ are elements in the ring of almost holomorphic 
modular functions of $\Gamma$. The latter is generated by a finite number of 
holomorphic and non-holomorphic generators. The relevant ones are constructed 
from the genus zero and genus one sector, i.e. ultimatively from the solutions of 
the Picard-Fuchs equations. The covariant derivative closes 
on these generators by (rigid) special geometry. The holomorphic anomaly equation 
can then be algebraically integrated w.r.t. the non-holomorphic modular generators. 
This leaves a holomorphic modular ambiguity, which is fixed by the gap conditions 
at the conifold discriminant.              
 
In section \ref{localp2} we exemplify the formalism and show that 
the topological string on a local Calabi-Yau geometry, which is the canonical 
line bundle over $\P^2$, is completely and very efficiently solved 
by our bootstrap approach. We also show how the generators, which we can 
construct in all cases from the solutions of the Picard-Fuchs equations 
relate in this case to classical modular functions on the $\Gamma_0(3)
\subset {\rm SL}(2,\Z)$ curve. We solve the theory to genus 105 
and present some of the holomorphic data at conifold, large structure 
point and orbifold point.  

In sections \ref{localF0} and \ref{localF1}
we extend this formalism to 
multi moduli examples. We show for the canonical 
bundle over $\HirzeF_0$ and $\HirzeF_1$, which have two 
parameters, how the gap condition at the conifold is again 
sufficient to fix all boundary conditions. In these cases the 
unknowns in the holomorphic ambiguity grow in leading order 
with $(c g)^2$ much faster then in the one moduli case. 
However, this is compensated by the fact that gap 
condition holds for all normal directions to the 
conifold discriminant in the complex two dimensional 
moduli space.           

In section \ref{conclusion} we discuss
relations of the results to $N=2$ Seiberg-Witten theory 
and general matrix models for which the spectral curve 
is a family of Riemann surfaces with $g>0$ and to open 
string amplitudes.  

The appendix \ref{modularity} reviews the necessary 
facts from the theory of modular functions. We try  
to give well known mathematical concepts a physical 
interpretation, which might shed some light on the 
relation between the holomorphic and the modular 
anomaly.

\section{Local Mirror Symmetry}
\label{localmirrorsymmetry}
The term local mirror symmetry refers to mirror symmetry 
for non-compact Calabi-Yau manifolds. Examples for the 
$A$-model geometry are the canonical line bundle 
$\K_S={\cal O}(-K_S)\rightarrow S$ over a Fano surface~\footnote{~Simpler examples 
involve line bundles over a complex curve such as 
${\cal O}(2(g-2)+k)\oplus {\cal O}(-k)\rightarrow {\cal C}_g$~\cite{BP} or manifolds $M$, which are 
given by a toric tree diagrams of the degeneration locus that correspond to genus 
$0$ mirror curves.} $S$. 
The compact part of B-model geometry is in this case given 
by a family of elliptic curves and a meromorphic 
differential. Using toric geometry as below 
an infinite set of examples of non-compact three-folds can be constructed.
They have  a partial overlap with the  $\K_S$ 
cases namely $S=\P^1\times\P^1$ or $S=\P^2$ and blow-ups 
thereof $S=\B\P^2_1,\B\P^2_2$,$\B\P^2_3$. 
The mirror geometry are Riemann surfaces with a meromorphic differential, whose genus is given by the 
number of closed meshes in the degeneration locus in the base 
of symplectic fibration, where two $S^1$'s degenerate.          
For early applications of local mirror symmetry to BPS state counting and geometric 
engineering of gauge theories see~\cite{Klemm:1996hh} 
and \cite{Katz:1996fh} respectively. For a systematic formulation see~\cite{Chiang:1999tz}\cite{Hori:2000kt}\cite{Hori:2003ic}. 
Below we give a very short review 
of the techniques.

\subsection{The local A-model} 

The A-model geometry of a non-compact toric variety is given by a  
quotient 
\begin{equation} 
M=(\C^{k+3}-Z)/G,
\label{toricmanifold}
\end{equation} 
where $G=(\C^*)^k$~\cite{cox}. 
On the homogeneous coordinates $x_i\in \C$ the group $G$ acts 
like $x_i\rightarrow \mu_\alpha^{Q_i^\alpha} x_i$, $\alpha=1,\ldots, k$  
with $\mu_\alpha \in \C^*$, $Q_i^\alpha\in \Z$. Here $Z$ is the 
Stanley-Reisner ideal, which has to be chosen so that the above quotient $M$ exists as a 
variety\footnote{~We assume that $M$ is simplicial, or that a simplicial subdivision 
in coordinate patches exists.}. The standard example is $\P^n=
(\C^{n+1}-\{0\})/(\C^*)$, with $Q^1_i=1$, $i=1,\ldots, n$. We denote 
generically by ${\cal S}$ the compact part of $M$.  

As explained in~\cite{wittenphases} $M$ can also be viewed as the vacuum field configuration of a 
2d gauged linear $(2,2)$ supersymmetric $\sigma$ model. The coordinates $x_i\in \C$, 
$i=1,\ldots,k+3$ are the vacuum expectation values of chiral superfields transforming 
as  $x_i\rightarrow e^{i Q_i^\alpha  \epsilon_\alpha} x_i$, $Q_i^\alpha\in \Z$, 
$\epsilon_\alpha\in \R$, $\alpha=1,\ldots, k$ under the gauge group $U(1)^k$. The vacuum
field configuration are the equivalence classes under the gauge group, 
which fulfill in addition the $D$-term constraints 
\begin{equation} 
D^\alpha=\sum_{i=1}^{k+3} Q_i^\alpha|x_i|^2=r^\alpha, \quad \alpha=1,\ldots, k \ .
\label{dterm} 
\end{equation}
The $r^\alpha$ are the K\"ahler parameters $r^\alpha=\int_{C_\alpha} \omega$, where 
$\omega$ is the K\"ahler form and $C_{\alpha}$ are curves spanning the Mori cone, which is 
dual to the K\"ahler cone.  $r^\alpha\in \mathbb{R}_+$ defines the K\"ahler cone. 
For $M$ to be well defined, field configurations for which the dimensionality of the 
gauge orbits drop have to be excluded. This corresponds to the choice of $Z$. 
In string theory $r^\alpha$ is complexified to $T^\alpha=r^\alpha+i\theta^\alpha$ with 
$\theta^\alpha=\int_{C_\alpha} B$, where $B$ is the NS $B$-field, while in the gauged linear 
$\sigma$-model the $\theta^\alpha$ are the $\theta$-angles of the $U(1)^k$ gauge group.

One can always describe $M$ by a completely triangulated fan. In this case the $Q_i^\alpha$ 
are linear relations between the points spanning the fan. A basis of such relations, 
which corresponds to a Mori cone can be constructed from a complete triangulation 
of the fan. $Z$ likewise follows combinatorially from the triangulation, see the 
examples\footnote{~Often there are many possible triangulation, which correspond to different
phases of the model~\cite{wittenphases}\cite{Aspinwall:1994zd}, e.g. K\"ahler cones 
connected by flopping a $\P^1$. The union of the cones define by all triangulations 
is called the secondary fan.}.

The Calabi-Yau condition $c_1(TM)=0$ holds if and only if\footnote{~Physically these are the conditions that the chiral $U(1)_A$ 
anomaly cancels in the gauged linear $\sigma$-model~\cite{wittenphases}.}   
\begin{equation}  
\sum_{i=1}^{k+3} Q_i^\alpha=0, \qquad \alpha = 1, \ldots, k .
\label{c1=0}
\end{equation}  
Note from \eqref{dterm} that negative $Q_i$ lead to non-compact directions in $M$, 
so that by (\ref{c1=0}) all toric Calabi-Yau manifolds $M$ are 
necessarily non-compact. To summarize, toric non-compact $A$-model 
geometries will be defined by suitably chosen charge vectors $Q_i^\alpha\in \Z$. 

We now come to invariants calculated by the $A$-model amplitudes. 
We consider maps $f:{\cal C}_g\rightarrow M$ from 
a genus $g$ curve ${\cal C}_g$, 
whose image curve is in the class $\beta\in H_2({\cal S},\Z)$. Now let 
as in~\cite{Klemm:1999gm}
\begin{equation} 
r_\beta^g=\int_{\overline {\cal M}(\beta,{\cal S})} c_{vir}(U_\beta) \ ,
\end{equation}  
with $U_\beta$ the bundle whose fiber over $({\cal C},f)\in \overline{ 
{\cal M}(\beta,{\cal S})}$ is $H^1({\cal C}_g,f^*M)$, be the Gromov-Witten invariant.   
The classical task in the closed topological $A$-model is to calculate the 
generating function
\begin{equation}
\cF=\log({\cal Z})=\sum_{g=0}^\infty \lambda^{2g-2} {\cal F}_g(Q)={c(T)\over \lambda^2}+l(T)+ \sum_{g=0}^\infty \sum_{\beta} 
\lambda^{2g-2} r_\beta^g Q^\beta\ , 
\end{equation} 
with $Q^\beta=\exp(2 \pi i \sum_{i=1}^{b_2(S)} \beta_i T_i)$, $\beta_i\in \Z_+$,
involving all closed string Gromov-Witten invariants as well as classical intersection numbers 
of the harmonic $(1,1)$-forms  $\frac{1}{3!}T^\alpha T^\beta T^\gamma \int_{M} \omega_\alpha\wedge 
\omega_\beta \wedge \omega_\gamma$ in the cubic $c(T)$ and $\frac{1}{24} T^\alpha \int_{M} c_2\wedge \omega_\alpha$ 
in the linear $l(T)$ term. The generating function $\cF$ can be reexpressed as one  
\begin{equation}
\cF={c(T)\over \lambda^2}+l(T)+\sum_{g=0}^\infty \sum_{\beta\in H_2(S,\Z)}\sum_{m=1}^\infty n_{\beta}^g {1\over m}
\left(2 \sin {m \lambda \over 2}\right)^{2 g-2}  Q^{\beta m}
\label{schwingerloop}
\end{equation}
for the BPS or Gopakumar-Vafa invariants $n_\beta^g\in\Z$ or with $q_\lambda=e^{i\lambda}$ the holomorphic 
partition function
\begin{equation}
\cZ=\sum_{\beta,k\in {\Z}} {\tilde n}^k_{\beta} 
(-q_\lambda)^k Q^\beta=\prod_{\beta}\left[\left(\prod_{r=1}^\infty (1-q_\lambda^r q^\beta)^{r n_\beta^0}\right)
\prod_{g=1}^\infty \prod_{l=0}^{2g-2}(1-q_\lambda^{ g-l-1} Q^\beta)^{(-1)^{g+r} 
\left(2 g-2\atop l\right) n_\beta^g}\right]\  
\label{zhol}
\end{equation}
becomes the generating function for the Donaldson-Thomas invariants\footnote{~Here we dropped the classical terms.} 
${\tilde n}^k_{\beta}\in \Z$.

\subsection{The local B-model} \label{bmodel}  

In the following we will describe the non-compact 
mirror $W$ following~\cite{Hori:2000kt,Katz:1996fh, Batyrev}.
Let $w^+, w^- \in \C$ and $x_i=:e^{y_i} \in \C^*$, $i=1,\ldots, k+3$ are 
homogeneous coordinates\footnote{~The $x_i$ here should not be identified with 
the $x_i$, which describe the $A$ model in the previous section.}, i.e. equivalence classes 
subject to the $\C^*$ action  
\begin{equation} \label{c*action}
x_i \mapsto \lambda x_i, \quad i=1,\ldots,k+3, \quad \lambda\in \C^*\ .
\end{equation} 
The mirror $W$ is defined from the charge vectors $Q_i^\alpha$ by the exponentiated $D$-term constraints
\begin{equation} 
(-1)^{Q_0^\alpha}\prod_{i=1}^{k+3} x_i^{Q_i^\alpha}=z_\alpha,\quad \alpha=1,\ldots, k \ . 
\label{exponentiateddterm}
\end{equation} 
and the general equation
\begin{equation} 
w^+ w^- =H=\sum_{i=1}^{k+3} x_i \ .
\label{mirrorgeometry} 
\end{equation}    
The Calabi-Yau condition (\ref{c1=0}) ensures the compatibility of (\ref{exponentiateddterm}) with 
(\ref{c*action}). Using the latter two equations to eliminate variables $x_i$  
in (\ref{mirrorgeometry}) $H$ can be parameterized by two variables 
$x=\exp(u),y=\exp(v) \in \C^*$ and the defining 
equations of the mirror geometry $W$ becomes 
\begin{equation} 
\label{mirrorcurve}
w^+ w^-=H(x,y; z_\alpha),
\end{equation}
which is a conic bundle over $\C^* \times \C^*$, where the conic fiber degenerates 
to two lines over the family of Riemann surfaces with punctures
\begin{equation} 
\S(z):=\{H(x,y; z^\alpha)=0 \} \subset \C^* \times \C^*\ ,
\end{equation} 
parameterized by the complex parameters $z^\alpha$. To establish that $W$ is a non-compact Calabi-Yau manifold 
note that 
\begin{equation}
\Omega={d H d x  d y\over H x y}
\end{equation}
is a regularizable no-where vanishing holomorphic volume form on $W$. Its periods are 
regularizable in the sense that $H,y$ can be integrated out to yield a meromorphic 
one-form on $\S$ 
\begin{equation}
\lambda={{\rm log}(y) \dd x \over x} \ ,
\end{equation}
whose periods clearly exist. They  are annihilated by the 
linear differential operators
\begin{equation}
D_\alpha=\prod_{Q_i^\alpha>0} \partial_{x_i}^{Q_i^\alpha} -
\prod_{Q_i^\alpha<0} \partial_{x_i}^{-Q_i^\alpha} \ .
\label{picardfuchs}
\end{equation}    
The redundancy in the parameterization of the complex structure is 
removed using the relations (\ref{exponentiateddterm}) and the scaling relation (\ref{c*action}). 
To do that it is convenient to write the differential operator
(\ref{picardfuchs}) in terms of logarithmic derivatives $\theta_i:=x_i
\partial_{x_i}$ and transform to logarithmic derivatives $\Theta_\alpha:=z_\alpha
\partial_{z_\alpha}$ using $\theta_i=Q_i^{\alpha} \Theta_{\alpha}$. 

As it is well known the solutions to (\ref{picardfuchs}) 
are constructed by the Frobenius method~\cite{Chiang:1999tz}, i.e. defining  
\begin{equation}
w_0({\underline{z}},{\underline{\rho}})=
\sum_{{\underline n}^{\underline{\alpha}}} \frac{1}
{\prod_{i} \Gamma[Q^\alpha_i (n^\alpha+\rho^\alpha)+1]} ((-1)^{Q^\alpha_0} 
z^\alpha)^{n^\alpha},
\end{equation}
then 
\be
X^0=w_0({\underline{z}},{\underline{0}})=1, \qquad {T}^\alpha=\frac{\partial }{2 \pi i \partial \rho^\alpha}
w_0({\underline{z}},{\underline{\rho}})|_{\underline{\rho}=0}
\label{gensol1}
\ee
are solutions. Note that the flat coordinates $T^\alpha$ approximate $T^\alpha\sim \log(z^\alpha)$ in 
the limit $z^\alpha\rightarrow 0$. Higher derivatives 
\be
X^{(\alpha_{i_1}\ldots \alpha_{i_n})}=\frac{1}{(2 \pi i)^n} 
\frac{\partial }{\partial \rho^{\alpha_{i_1}}}
\ldots\frac{\partial }{\partial \rho^{\alpha_{i_n}}}
w_0({\underline{z}},{\underline{\rho}})|_{\underline{\rho}=0}
\label{gensol2}
\ee
 also obey the 
recursion imposed by (\ref{picardfuchs}), \emph{i.e.} they fulfill 
(\ref{picardfuchs}) up to finitely many terms. However, a unique, up to addition 
of previous solutions, linear combinations of the $X^{\alpha_{i_1}\ldots \alpha_{i_2}}$ 
is actually the last solution of the Picard-Fuchs system. This solution encodes the genus zero
Gromov-Witten invariants. It is a derivative of the holomorphic prepotential ${\cal F}_0$ and 
the triple intersection $C_{ijk}=\partial_{T_i}\partial_{T_j}\partial_{T_k}{\cal F}_0 $ 
can be constructed from it, see the examples for more details. We will turn to generating 
functions for the higher genus amplitudes 
in the next section.

\section{Integrability of the holomorphic anomaly equation}
\label{directintegration}
This section is to review the recent results of~\cite{GKMW}\cite{Alim:2007qj} on the polynomial recursive solution of 
the holomorphic anomaly equation of~\cite{BCOV2} and to set our conventions. This recursive solution is a 
generalization of the pioneering work of Yamaguchi and Yau~\cite{Yamaguchi:2004bt} who observed that the 
non-holomorphic dependence of the topological free energy function of the quintic can be expressed by a 
finite number of generators. Our main focus is the local geometry, hence we will mainly explain how 
the formalism simplifies in the non-compact case.

\subsection{Direct Integration in local Calabi-Yau geometries}
One of the main tasks in topological string theory is to compute the free energies $F_g$ appearing in the 
topological string partition function $Z=\exp(\sum\lambda^{2g-2}F_g)$. We will assume that the genus 
zero sector has been determined from the solutions to the Picard-Fuchs equations discussed  
in section \ref{bmodel}. The genus one amplitude is 
associated to the Ray-Singer torsion of the Calabi-Yau space~\cite{BCOV2}. It fulfills 
a special holomorphic anomaly equation, which is integrated to~\cite{BCOV1}\footnote{~ In the following we denote 
the non-holomorphic quantities by straight characters $F_g$ and the holomorphic 
limits by calligraphic characters ${\cal  F}^f_1$, with a label $f$ of the patch, 
where the limit is taken.}
\begin{equation}
F_1={1\over 2}\log\left[\exp \left[K\left(3+h^{1,1}-\frac{\chi}{12}\right)\right] 
\det G_{i\bar \jmath}^{-1} |f_1|^2\right].
\label{1F1}
\end{equation}
While the exponential of the real K\"ahler potential $\exp(K)\sim X^{0}\rightarrow 1$
in the holomorphic limit in the non-compact models~\cite{Klemm:1999gm}, 
the $F_1$ is non-holomorphic due to the K\"ahler  metric $G_{i\bar \jmath}$ on the complex structure moduli space. 
$f_1$ is the holomorphic ambiguity in this integration and it can be argued to 
be a power of the discriminant loci of $\S$~\cite{BCOV1,Ghoshal:1995wm},
i.e. $f=\prod_{i}\Delta_i^{a_i}\prod_{i=1}^{h^{2,1}}z_i^{b_i}$. 
The parameters, $a_i,b_i$, can be solved from the limiting behavior of $F_1$
near singularities,
$\lim_{z_i\rightarrow0}F_1=-\frac{1}{24}\sum_{i=1}^{h^{2,1}}t_i\int_M c_2J_i$ 
as well as the universal behavior at conifold singularities $a_{\rm con}=-\frac{1}{12}$.
 
As was shown in \cite{BCOV2} $F_g$ is for $g>1$  a non-holomorphic section of a line bundle $\cL^{2-2g}$ which 
fulfills a recursive differential equation
\begin{equation}
\dbar_\bi F_g=\frac{1}{2}\bar{C}_\bi^{jk}\left(D_jD_kF_{g-1}+\sum_{r=1}^{g-1}D_jF_{g-r} D_kF_r\right),\qquad (g>1)
\label{HAEq}
\end{equation}
called the holomorphic anomaly equation.
The covariant derivatives contain the connection $\partial_i K=K_i$ of $\cL$ and the Christoffel symbols $\Gamma^i_{jk}$ of the
K\"ahler metric. The recursive nature is due to the fact that Riemann surfaces with marked points split 
at the boundary of moduli space, $\cM$, into either pairs of lower genus surfaces or surfaces with fewer marked points.

The key input for the direct integration procedure is the special geometry integration condition
\begin{equation} 
\bar \partial_{\bar \imath} \Gamma^k_{ij}=\delta_{i}^k G_{j\bar
  \imath}+\delta_{j}^k G_{i\bar \imath}- C_{ijl} \bar{C}^{kl}_{\bar \imath}\ .
\label{sk} 
\end{equation} 
Here $C_{ijl}$ are the holomorphic Yukawa couplings which transform as ${\rm Sym}^3(T \cM)\otimes \cL^{-2}$ and 
$\bar{C}^{kl}_{\bar \imath}=e^{2K} G^{k\bar k} G^{l\bar l} {\bar C}_{\bar \imath\bar k \bar l}$.

(\ref{sk}) implies that the propagator $S^{ij}$, which is defined by 
$\bar \partial_{\bar  k} S^{ij}=\bar{C}^{ij}_{\bar k}$, can be solved from 
the integrated version of (\ref{sk}) \cite{BCOV2} 
\begin{equation} \label{e:propeq} 
\Gamma^k_{ij}=\delta_{i}^k \partial_j K+\delta_{j}^k \partial_i K- C_{ijl}
S^{kl}+\tilde f_{ij}^k\ , 
\end{equation}                  
up to the holomorphic ambiguity $\tilde f^k_{ij}$. Taking the anti holomorphic
derivative, using (\ref{sk}) and  $\partial_{\bar \jmath} S^k=S_{\bar
  \jmath}^k$ 
it follows that
\begin{equation}
{\bar \partial}_{\bar k}(D_i S^{kl}) ={\bar \partial}_{\bar k} (  \delta_{i}^k
S^l+\delta_{i}^l S^k - C_{inm} S^{km} S^{ln})\  ,
\end{equation} 
and so 
\begin{equation}
D_i S^{kl} =  \delta_{i}^k S^l+\delta_{i}^l S^k - C_{inm} S^{km} S^{ln}+
f^{kl}_i\ .
\end{equation}

In the local case one has the following simplifications\footnote{~In the global
case on needs further the closing of covariant derivatives of $S^i$ and $S$
with $\partial_{\bar \imath} S=G_{\bar \imath j}S^j$. This has been discussed
in~\cite{Yamaguchi:2004bt},\cite{GKMW} and particular nicely in~\cite{Alim:2007qj}.}. 
The K\"ahler connection in $D_i$ becomes trivial, and the $S^l$, 
(as well as the $S$, see \cite{BCOV2}) vanish, \emph{i.e.} the above equation
becomes simply         
\begin{equation} \label{DS}
D_i S^{kl} = -C_{inm} S^{km} S^{ln}+ f^{kl}_i .
\end{equation} 
Also, the K\"ahler connection $\partial_j K$ in (\ref{e:propeq}) drops out, so
the $S^{ij}$ are solved from 
\begin{equation} \label{PropEq}
\Gamma^k_{ij}=- C_{ijl} S^{kl}+\tilde f_{ij}^k\ .
\end{equation}  
Note that this is an over-determined system in the multi moduli case which
requires a suitable choice of the ambiguity $\tilde f_{ij}^k$. This choice is 
simplified by the fact~\cite{Aganagic:2002wv} that $\d_i F_1$ can be expressed through the 
propagator as
\begin{equation}
\d_i F_1=\half C_{ijk}S^{jk}+A_i,
\label{F1prop}
\end{equation}
with an ambiguity $A_i$, which can be determined by the ansatz $A_i=\d_i(\tilde{a}_j\log\Delta_j+\tilde{b}_j\log z_j)$.

Once the $S^{ij}$ are obtained and the ambiguities in (\ref{DS},\ref{PropEq}) 
have been fixed, the direct integration  of (\ref{HAEq}) is quite simple. Everything on the right 
hand side of the holomorphic anomaly equation (\ref{HAEq}) can be rewritten in terms of the generators $S^{ij}$ 
and holomorphic functions. 
If we further express the anti-holomorphic derivative of $F_g$ as
\begin{equation}
\dbar_\bi F_g=\bar{C}_\bi^{jk}\frac{\d F_g}{\d S^{jk}},
\end{equation}
and assume linear independence of $\bar{C}_\bi^{jk}$, (\ref{HAEq}) can be written as
\begin{equation}
\frac{\d F_g}{\d S^{jk}}=\frac{1}{2}(D_j\d_kF_{g-1}+\sum_{r=1}^{g-1}\d_jF_{g-r}\d_kF_r).
\label{holomorphicanomaly}
\end{equation}
This equation can easily be integrated w.r.t. $S^{ij}$  and it can be shown that $F_g$ is a
polynomial in $S^{jk}$ of degree $3g-3$.

\subsection{Fixing the ambiguity}
Due to the equation (\ref{holomorphicanomaly}) the iteration in the genus
is in principle quite easy on the B-model side and the topological invariants
of the A-model geometry can be extracted without effort. However, the issue is
fixing the holomorphic ambiguity $f_g$ arising after each integration
step w.r.t. the $S^{ij}$. Modularity, regularity at the orbifold point and at 
the large radius point, as well as the leading behavior at the conifold 
singularities~\cite{Ghoshal:1995wm} imply the following ansatz for $f_g$ 
\begin{equation}
f_g=\sum_{i}\frac{A^i_g}{\Delta_i^{2g-2}},
\end{equation}
where $A^i_g$ is a polynomial in $z$ of degree $(2g-2)\cdot\deg\Delta_i$
and the sum runs over all irreducible components of the discriminant locus. 
Note that the moduli space ${\cal M}(\S)$ allows a compactification, 
which includes only the ordinary double point discriminants or conifolds at 
complex codimension one loci in the moduli space.     
$A^i_g$ are polynomials in the monodromy invariant variables $z_i$,
$i=1,\dots, n$ of the model. Their degree is bounded by regularity of 
the $F_g$ in the  limit that these variables tend to infinity by the degree 
of the $\Delta_i$. In general this implies a growth of the unknowns 
roughly with $(c_i g)^n$, where $c_i$ depends on the degrees of 
$\Delta_i$. However, if we approach a conifold singularity we 
also find in the multi parameter case a gap. It is of the form 
\begin{equation}  
\F^c_g= \frac{c^{g-1} B_{2g}}{2g (2g-2)t_c^{2g-2}}+{\cal O }(t_c^0) \ . 
\label{gapm}
\end{equation}  
where we approach a conifold in the limit $t_c\rightarrow 0$, with 
$t_c$ a flat coordinate normal to the singularity\footnote{~$c$ is an 
undetermined constant, which can be absorbed by rescaling the variable
$t_c$.} (see Figure 3). The coefficients of the sub-leading powers of $t_c$ depend generically on the further 
$n-1$ directions, which are tangential to the discriminant locus. For 
a generic choice of coordinates these coefficients are (infinite) series 
in the tangential $n-1$ variables. However, demanding the vanishing of these 
coefficients is an over-determined system and it is not easy to count the 
independent conditions. But in local models where the geometry of the B-model 
is completely encoded in a Riemann surface of genus $g>0$ we find that 
the gap condition is  sufficient to determine all parameters in the 
ambiguity except for the one, which corresponds to the constant term in $F_g$. 
The latter can be determined by the known constant map contribution to $\cF_g$ 
at the point of large radius in moduli space
\begin{equation}
\F_g=\frac{\chi B_{ 2 g-2} B_{2g}}{4 g (2 g-2) (2 g-2)!}+\cO(Q).
\end{equation}
Therefore we find that the holomorphic anomaly equations are completely 
integrable for local Calabi-Yau spaces. Our claim that this is true 
in general is motivated by the fact that the only type of degeneration of a
Riemann surface in complex codimension one is the nodal degeneration and the 
leading local behavior of the $F_g$ at this singularity is always governed by 
the gap structure and in particular the argument for the existence of the
gap~\cite{HKQ} does not depend on the direction nor on the particular point at 
which the conifold locus is approached.

\section{$\K_{\P^2}=\cO(-3)\rightarrow\P^2$}
\label{localp2}

The toric data of $\K_{\P^2}$ is summarized in the following matrix 
\begin{equation}
(V|Q)=\left(\begin{array}{ccc|c} 0&0&1&-3\\1&0&1&1\\ 0&1&1&1\\ -1&-1&1&1\end{array}\right)\ 
\end{equation}
The $A$-model is described from these data as follows. The generators of the
toric fan
$\mathbb{F}$  $v_i$, $i=0,\ldots ,3$ are the rows of
$V$, while the columns of $Q$ are the charge vectors, which are the
coefficients of linear relations among the $v_i$. To each $v_i$ we associate
homogeneous coordinates  $x_i$.
There is an unique complete triangulation of $\mathbb{F}$ into simplexes given
by ${\cal T} =\{\{v_0,v_1,v_2\},\{v_0,v_1,v_3\},\{v_0,v_2,v_3\}\}$. 
The Stanley-Reisner ideal $Z$ is generated by intersection of divisors
$D_i=\{x_i=0\}$, whose associated points are not on a common simplex in
${\cal T}$, i.e. by  $Z=\{x_1=x_2=x_3=0\}$.  The  $(x_1:x_2:x_3)$ are hence
the homogeneous coordinates of $\P^2$. 
The three $\C^3$ patches that cover the 3-fold 
$\K_{\P^2}$ are specified by the scaling in 
(\ref{toricmanifold}) as $(l_1=x_0 x_1^3;u_1=x_2/x_1,v_1=x_3/x_1)$, 
$(l_2=x_0 x_2^3;u_2=x_1/x_2,v_2=x_3/x_2)$ and $(l_3=x_0
x_3^3;u_3=x_1/x_3,v_3=x_2/x_3)$ with the obvious transition 
functions. 

The $B$-model geometry is defined by the one parameter family of Riemann
surfaces $\S(z)$
\begin{equation} 
H(x,y;z)=x + 1-z \frac{x^3}{y}+y=0\ .
\label{Bgeomp2}
\end{equation} 
Here we set  $x_1=1$ in (\ref{mirrorgeometry}) by the scaling relation
(\ref{c*action}) 
and eliminated $x_2$ using (\ref{exponentiateddterm}) in favor of $x:=x_0$ and $y:=x_3$. 

\subsection{Global properties of the moduli space of $\S(z)$} 
After writing (\ref{Bgeomp2}) in Weierstrass form in $\P^2$ 
we find the $j$-function of the elliptic family $\S(z)$ 
\begin{equation} 
j=-\frac{{\left( 1 + 24\,z \right) }^3}
    {z^3\,\left( 1 + 27\,z \right) } .
\label{j1}
\end{equation} 
Its  moduli space for the complex structure parameter $z$ is 
${\cal M}(\S(z))=\P^1\setminus \{z=0,z=-\frac{1}{27},\frac{1}{z}=0\}$. 
The critical points of $j$ are referred to as large radius point, conifold 
points and orbifold point,\footnote{~By using a multi covering  
variable $\psi=-\frac{1}{3 z^\frac{1}{3}}$ one gets three symmetric conifold points 
at $\psi^3=1$ and no orbifold point.} respectively.  

Following the description after  (\ref{picardfuchs}) we find
\begin{equation} 
{\cal D}=\Theta^3+ 3 z (3 \Theta-2)(3 \Theta-1)\Theta ={\cal L}\Theta ,
\label{pflocalp2}
\end{equation}
here ${\cal L}$ is the Picard-Fuchs equation for the periods over the holomorphic
differential $\omega=\frac{\dd x}{y}$. From this follows that 
\begin{equation}
z \frac{\dd}{\dd z} \lambda= \omega + exact, 
\label{ldw}
\end{equation} 
where $\lambda$  is the meromorphic differential. This meromorphic differential 
$\lambda$ has a pole with non-vanishing residue and we denote the cycle around 
this pole $\gamma$, while  $a,b\in H_1(\S,\Z)$ are 
a basis  for the integral  cycles on $\S$.  
On  $\hat \Pi=(\int_{b} \lambda,\int_{a} \lambda,\int_{\gamma} \lambda)^T$ the monodromy acts by 
\begin{equation}
M_{z=0}=\left(\begin{array}{ccc} 
1 \ & 3\ & 0 \\ 
0 \ & 1\ & 1 \\ 
1 \ & 0\ & 1 \end{array}\right),\quad M_{z=-\frac{1}{27}}=\left(\begin{array}{ccc} 
1 \ & 0\ & 3 \\ 
-1 \ & 1\ & 1 \\ 
0 \ & 0\ & 1 \end{array}\right),\quad M_{\frac{1}{z}=0}=M_{z=-\frac{1}{27}}^{-1} M_{z=0}^{-1}\ ,
\end{equation} 
as can be seen explicitly by analytic continuation of the periods into the three patches 
near the singular points (\ref{periodslargeradius},\ref{periodsconifold}) 
as well as (\ref{periodsorbifold},\ref{analyticlargeorb}). It follows from the  
monodromy invariance of $z$ and (\ref{ldw}), that the upper left $(2\times 2)$ block in the 
above matrices acting on $\hat \Pi$ represents also the monodromy action 
on the $\Pi=(\int_{b} \omega,\int_{a} \omega )^T$. The later generates
\begin{equation} 
\Gamma^0(3)=\left\{\left(\begin{array}{cc} a \ & b \\ c &
      d\end{array}\right)\in {\rm SL}(2,\Z)\,\Biggr|\, b\equiv0\ {\rm mod}\ 3\right\}\ .
\end{equation}

\begin{figure}
\center
\includegraphics[width=8cm]{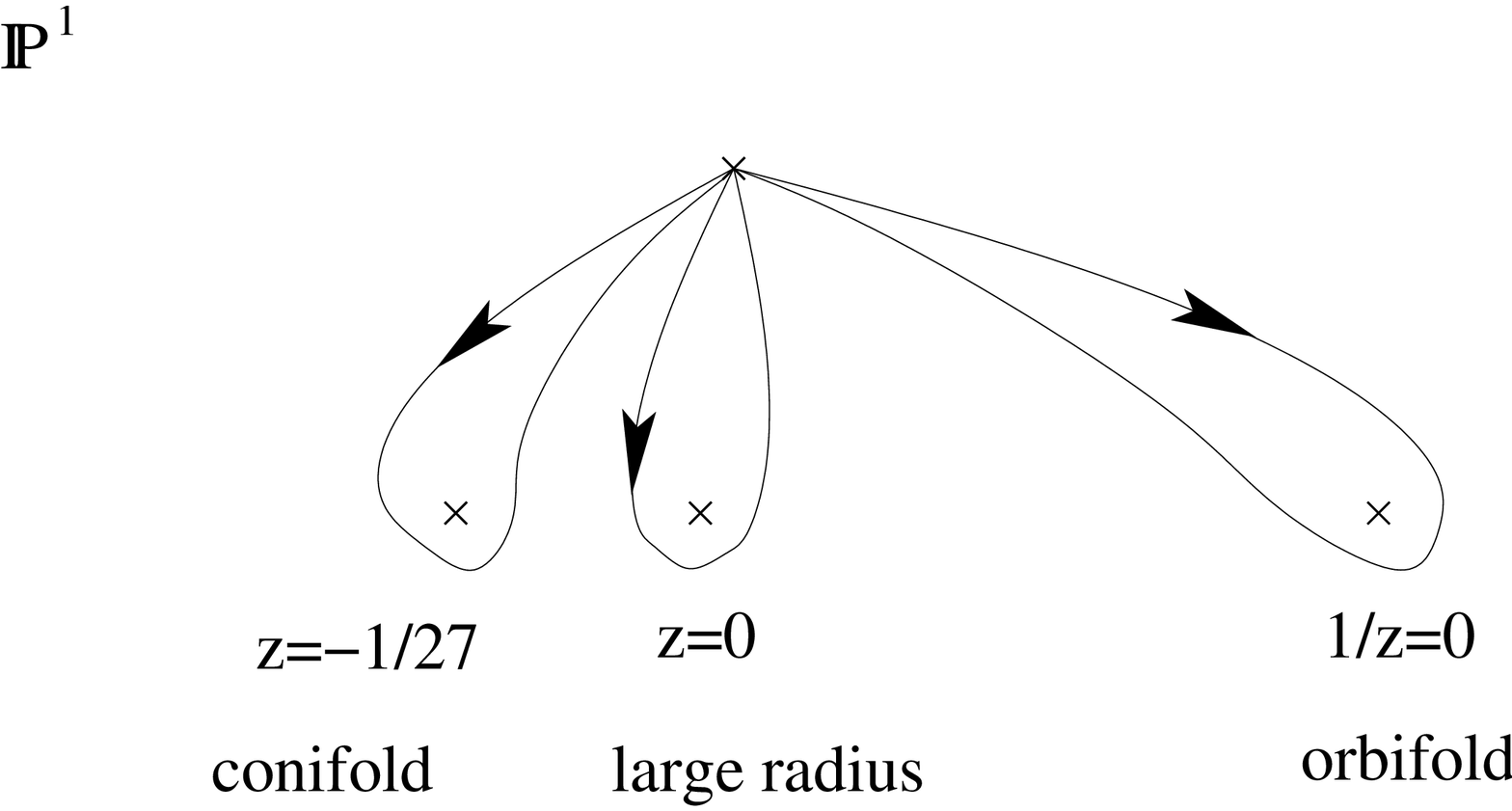}
\caption{Definition of the monodromies in ${\cal M}(\S(z))=\P^1\setminus \{z=0,z=-\frac{1}{27},\frac{1}{z}=0\}$.}
\label{mon}
\end{figure}

\subsection{Periods and genus zero and one amplitudes in all patches} 

We review now the construction of the holomorphic prepotential encoding 
the genus zero amplitude and the an-holomorphic Ray-Singer torsion 
encoding the genus one amplitude in the patches near the three singular points 
described above. In each patch we introduce appropriate flat coordinates, distinguished
by the monodromies around the critical points.  Once the 
flat coordinate is chosen one can consider  a holomorphic limit 
of the amplitudes  for $g>0$. This yields  holomorphic generating functions for 
certain topological invariants, depending on the point in moduli
space. Notably the Gromov-Witten invariants near $z=0$ 
and the orbifold Gromov-Witten invariants near $\frac{1}{z}=0$. 
The most useful structure for the integrability comes from the gap 
in the expansion at the conifold. 

\subsubsection{Expansion near the large radius point} 

The solutions near $z=0$ are according to 
(\ref{gensol1},\ref{gensol2}) given\footnote{~We also note that the system (\ref{pflocalp2}) is related to the Meijer
G-functions and $T=-\frac{1}{2 \pi i\Gamma\left(\frac{1}{3}\right)\Gamma\left(\frac{2}{3}\right)}G^{3\,3}_{2\,2}
\left({{\frac{1}{3}\ \frac{2}{3} \ 1} \atop {0 \ 0 \ 0}}\biggr| 27 z\right)$.} by $\omega_0(z,0)=1$, 
$X^{(1)}={1\over 2 \pi i}(\log(z) + \sigma_1(z))$ and $X^{(1,1)}=
{1\over (2 \pi i)^2} (\log(z)^2+2 \sigma_1\log(z) + \sigma_2(z))$, 
where the first orders are $\sigma_1=-6\,z + 45\,z^2 +{\cal O}(z^3)$ 
and $\sigma_2=-18\, z + \frac{423\, z^2}{2} +{\cal O}(\tilde z^3)$.
The actual integral basis of periods is given by the linear combinations
\begin{equation}
\hat \Pi=
\left(\begin{array}{c}
T_D\\
T\\
1\end{array}\right) 
=\left(\begin{array}{c}
-9 \partial_{T} {\cal F}_0\\
T\\
1\end{array}\right)
 =\left(\begin{array}{c}
{3\over 2} X^{(1,1)}-{3\over 2} T+{3\over 4}\\
X^{(1)}\\
1\\
\end{array}\right)
\label{periodslargeradius}
\end{equation}

In order to express ${\cal F}_0$ in terms of the flat coordinate 
$T$, we introduce  the monodromy invariant quantity $Q=e^{2\pi iT}$ 
and invert. This yields the large radius mirror map
\begin{equation}
z(Q)=Q + 6\,Q^2 + 9\,Q^3 + 56\,Q^4 - 300\,Q^5 + \ldots
\end{equation}

The normalization $T_D=-9 \partial_T \F_0$ is such that $\F_0$ is the generating 
function for the genus zero Gromov-Witten invariants of ${\cal
  O}(-3)\rightarrow \P^2$ in the normalization that reproduces the $A$-model 
results obtained first by localization~\cite{Klemm:1999gm}, see table B.1 
for the BPS invariants
\begin{equation}
{\F}_0=-\frac{{T}^3}{18} +\frac{T^2}{12}-\frac{T}{12} + 3\,Q - 
  \frac{45\,Q^2}{8} + \frac{244\,Q^3}{9} - 
  \frac{12333\,Q^4}{64} + 
  \frac{211878\,Q^5}{125} + \ldots
\end{equation}
The normalization of the Yukawa coupling, with which we get this expansion
is 
\begin{equation}
C_{z z z}=-\frac{1}{3}\frac{1}{z^3 (1+27 z)}\ .
\label{yukawa1lp2}
\end{equation}
The Yukawa coupling transforms as ${\rm Sym}^3(T \cM)\otimes \cL^{-2}$, where
the K\"ahler connection, i.e. the line bundle $\cL$  is trivial in the local 
case. From the special K\"ahler relations in flat coordinates we get
\begin{equation}
\left(\frac{\partial }{\partial T}\right)^3{\F}_0= 
C_{TTT}= \left(\frac{\partial z} {\partial T}\right)^3 C_{z z z}\ .
\label{yukawa2lp2}
\end{equation}
Note that (\ref{yukawa1lp2}) is modular invariant and valid in all ${\cal M}(\S)$. 
The expression (\ref{yukawa2lp2}) on the other hand requires a choice of the 
flat coordinate $T$, which is only canonical near $z=0$. One can view $T$ as the 
coordinate and $P_T=\partial_T {\cal F}_0$ as the dual momentum and show 
that $Z=\exp(F)$ transforms as a wavefunction under a change of 
polarization, i.e. when a different choice (related by a linear 
transformation) for coordinates and  momenta is 
made~\cite{Witten:1993ed}\cite{ABK}.

Using the standard definition of the modular parameter of the family of 
elliptic curves $\tau=\frac{\int_b \omega}{\int_a \omega}$, (\ref{ldw}) and
(\ref{periodslargeradius}) we find 
\begin{equation}
\tau=\frac{\frac{\partial T_D}{\partial z}}{\frac{\partial T}{\partial z}}=
-9 \frac{\partial^2 \F_0}{\partial^2 T}\ .
\label{j3}
\end{equation}
The resulting relation $z(q)$, with
$q=\exp(2 \pi i \tau)$ has to be compatible with (\ref{j1}). Indeed inserting $z(q)$ into (\ref{j1})
yields the standard expansion of the elliptic $j$-function (\ref{jfunct}).
Using $z(q)$ we can express the non-holomorphic genus one potential as 
\begin{equation}
F_1 =-\log\left(\sqrt{\tau_2} \eta(q) \bar \eta(\bar q)\right) 
-  \frac{1}{24}\log\left(1+\frac{1}{27 z}\right)\ .
\label{2F1tau}
\end{equation}
Both the Dedekind $\eta$ function as well as  $1+\frac{1}{27 z}$ are
powers of the discriminant of $\S$. The former transforms with  weight $\frac{1}{2}$
that is canceled by that of $\tau_2$ (\ref{imtautrans}). We note that both forms of $F_1$ 
(\ref{1F1}) and  (\ref{2F1tau}) are manifestly modular invariant.

Using $\det G_{i\bar \jmath}^{-1}\rightarrow C \det \frac{\partial z_i}
{\partial T_j}$ in the holomorphic limit $\bar T \rightarrow \infty$ or
equivalently  $\tau \rightarrow i\infty$ one gets up to irrelevant constants
the holomorphic expression
\begin{equation}
{\cal F}_1=\frac{1}{2} \log\left(\frac{\partial z}{\partial T}\right)-
\frac{1}{12}\log(z^7(1+ \frac{1}{27 z}))\ .
\label{f1}
\end{equation} 
This expression is not modular invariant and depends on the choice of our
special coordinate. It does give however the generating function 
for GW invariants at genus one 
\begin{equation}
{\cal F}_1=\frac{T}{12}+\frac{Q}{4} - 
  \frac{3\,{{Q}}^2}{8} - 
  \frac{23\,{{Q}}^3}{3} + 
  \frac{3437\,{{Q}}^4}{16} - 
  \frac{43107\,{{Q}}^5}{10} + \ldots
\end{equation}
in accordance with~\cite{Klemm:1999gm}, see table B.1 
for the BPS invariants. 
 
\subsubsection{Expansion near the conifold}

To obtain the closed variables at the conifold we solve the 
Picard-Fuchs equation after the variable transformation 
$z=\frac{\Delta-1}{27}$. The basis of periods at large 
radius (\ref{periodslargeradius}) has the following expansion 
at the conifold point
\begin{equation}
\Pi=\left(\begin{array}{c}
a \, t_c\\
3\, a\, {t_c}_D\\
1\end{array}\right) 
=\left(\begin{array}{c}
a\, t_c\\
3\, a\, \partial_{t_c}\ {\cal F}^{c}_0
\\
1\end{array}\right)
 =\left(\begin{array}{c}
 a(\Delta + \frac{11\,{\Delta}^2}{18} + \frac{109\,{\Delta}^3}{243} + {\cal O}(\Delta^4)) \\
a\left(a_0+a_1 t_c -\frac{1}{2 \pi i} (t_c \log(\Delta)+\frac{7\,{\Delta}^2}{12} + \frac{877\,{\Delta}^3}{1458} + {\cal O}(\Delta^4))\right)\\
1\\
\end{array}\right)\ ,
\label{periodsconifold}
\end{equation}
where $a=-\frac{\sqrt{3}}{2\pi }$, $a_0=-\frac{\pi}{3} -1.678699904 i=
\frac{1}{i\sqrt{3}\Gamma\left(\frac{1}{3}\right)\Gamma\left(\frac{2}{3}\right)}G^{3\,3}_{2\,2}
\left({{\frac{1}{3}\ \frac{2}{3} \ 1} \atop {0 \ 0 \ 0}}\biggr| -1\right)$ 
and
$a_1=\frac{3 \log(3)+1}{2 \pi i}$. 

The natural local flat coordinate at the conifold is $t_c$ and with the
conifold mirror map 
\begin{equation}
\Delta=t_c-\frac{11 t_c^2}{18}+\frac{145 t_c^3}{486}-\frac{6733
  t_c^4}{52488}+{\cal O}\left(t_c^5\right)
\end{equation}
the genus zero prepotential becomes 
\begin{equation} 
\label{conifoldgenus0}
{\cal F}^{c}_0=c_0 + \frac{a_0}{3} t_c+\left(\frac{a_1}{6}-\frac{1}{12}\right) t_c^2+ 
t_c^2 \frac{\log (t_c)}{6}-\frac{t_c^3}{324} + \frac{t_c^4}{69984} 
+ \frac{7\,t_c^5}{2361960} - \frac{529\,t_c^6}{1700611200} +{\cal O}(t_c^7)\ .
\end{equation}
Note that we rescaled $t_c$ by $a$ in order to avoid non 
rational numbers in this expansion and the extra factor $3$ in
(\ref{periodsconifold}) is so that $\partial_{t_c}^3 {\cal
  F}^{c}_0=\left(\frac{\partial z}{\partial t_c}\right)^3 C_{zzz}(t_c)$  
We can also find the holomorphic limit of the genus one prepotential as
\begin{equation}
{\cal F}^{c}_1=\frac{1}{2} \log\left(\frac{\partial z}{\partial t_c}\right)
-\frac{1}{12}\log(z^7(1+ \frac{1}{27 z}))\ 
\label{f1b}
\end{equation}
and expand it as

\begin{equation} 
\label{conifoldgenus1}
{\cal F}^{c}_1=c'_0 - \frac{\log (t_c)}{12}+
\frac{5\,t_c}{216} - \frac{t_c^2}{23328} - \frac{5\,t_c^3}{157464} + 
\frac{283\,t_c^4}{75582720} - \frac{43\,t_c^5}{153055008} + \frac{4517\,t_c^6}{385698620160}+{\cal O}(t_c^7)\ .
\end{equation}

\subsubsection{Coordinates and amplitudes at the Orbifold} 
\label{orbifold}
At the orbifold point, the model admits an exact field theory 
description as an orbifold of three complex bosons
$\C^3/\Z_3$.    
After transforming the Picard-Fuchs equation to the 
$\psi=-\frac{1}{3 z^{\frac{1}{3}}}$ coordinate we 
find  the following local expansion of a basis of 
solutions $(1,B_1,B_2)$ with
\begin{equation} 
B_k=(-1)^{\frac{k}{3}+k+1}\frac{(3 \psi)^k}{k} \sum_{n=0}\frac{\left[\frac{k}{3}\right]_n^3}
{\prod_{i=1}^3 \left[\frac{k+i}{3}\right]_n} \psi^{3n}\ ,
\end{equation}
where $[a]_n=a (a+1) \ldots (a+n+1)$ is the Pochhammer symbol. We define orbifold periods, which
diagonalize the $\Z_3$ orbifold monodromy action
\begin{equation}
\Pi_{orb} =
\left(\begin{array}{c}
\sigma_D\\
\sigma\\
1\end{array}\right)=\left(\begin{array}{c}
-3 \partial_\sigma \F^{orb}_0\\
\sigma\\
1\end{array}\right)
 =\left(\begin{array}{c}
B_2\\
B_1\\
1\\
\end{array}\right)\ ,
\label{periodsorbifold}
\end{equation}
i.e. $(B_2,B_1,1)\mapsto (\exp\left(4\pi i\over 3\right) B_2, \exp\left(2\pi
  i\over 3\right) B_2,1)$ under $\psi\mapsto \exp\left(2\pi i\over 3\right) \psi$.
Note, that this is not the basis at large radius, but rather connected to
it by the transformation $\Pi=M \Pi_{orb}$ with 
\begin{equation}
M=
\left(\begin{array}{rrr}
-\frac{3}{1-\alpha} A& \frac{3 \alpha}{1-\alpha} B& 1\\
A&  B& 0\\
0& 0& 1\end{array}\right)\ .
\label{analyticlargeorb}
\end{equation}
Here we introduced 
\begin{equation}
A:=\frac{i\Gamma\left(\frac{2}{3}\right)}{2 \pi 
  \Gamma^2\left(\frac{1}{3}\right)}\ , \quad 
B:= \frac{i
  \Gamma\left(\frac{1}{3}\right)}{2\pi \Gamma^2\left(\frac{2}{3}\right)}\ , \quad
\alpha:=\exp\left(\frac{2\pi i}{3}\right)\ .
\end{equation} 

We normalize the flat coordinate $\sigma$ and ${\cal F}_0^{orb}$  
to match the orbifold Gromov-Witten invariants of \cite{Coates} 
in the orbifold prepotential   
\begin{equation}
\label{orbifoldgenus0}
{\cal F}^{orb}_0=\frac{\sigma^3}{18} - \frac{\sigma^6}{19440} + 
  \frac{\sigma^9}{3265920} - 
  \frac{38497\,\sigma^{12}}{2571324134400} +\ldots
\end{equation}
and the special geometry relation  $\partial^3_\sigma {\cal F}^{orb}_0=
\left(\frac{\partial z}{\partial \sigma}\right)^3 C_{zzz}(\sigma)$, which
implies the orbifold mirror map
\begin{equation} 
\frac{\psi}{\alpha^2}=\frac{\sigma}{3}+\frac{\sigma^4}{1944}-\frac{29
    \sigma^7}{11022480}+{\cal O}(\sigma^{10})\ .
\end{equation}
The expansion for the holomorphic limit of the Ray-Singer Torsion reads   
\begin{equation}
{\cal F}^{orb}_1=\frac{1}{2} \log\left(\frac{\partial z}{\partial \sigma}\right)
-\frac{1}{12}\log(z^7(1+ \frac{1}{27 z}))=c_0+ \frac{\sigma^6}{174960} - \frac{\sigma^9}{6298560} + 
  \frac{13007\,\sigma^{12}}{3142729497600} + \ldots \ . 
\end{equation}

\subsection{Direct integration for $\K_{\P^2}$}

Let us now discuss the direct integration for the 
non-compact $\K_{\P^2}$ geometry. 
Here we have only one propagator, which we denote in the $z$ 
variables by $S^{zz}$. The propagator has a holomorphic ambiguity, 
which we may choose by imposing in (\ref{F1prop}) the vanishing of $A_z$
\begin{equation} 
S^{zz}=\frac{2}{C_{zzz}} \partial_z F_1\ .  
\label{prop}
\end{equation} 

This implies the following ambiguity factors in (\ref{e:propeq})
\begin{equation}
\Gamma^z_{zz }=- C_{zzz} S^{zz}-\frac{7 + 216 z}{6 z \Delta}
\label{GG} 
\end{equation}  
and in (\ref{DS})
\begin{equation}
D_z S^{zz} = - C_{zzz} S^{zz} S^{zz}-\frac{z}{12 \Delta} \ .
\label{DSS} 
\end{equation} 
The right hand side of equation (\ref{holomorphicanomaly}) is easily 
evaluated using the connection $\Gamma^z_{zz}$ and yields e.g. for $g=2$ using (\ref{prop}),
(\ref{GG}) and (\ref{DSS})     
\begin{equation} 
\partial_{S^{zz}} F_2  =C^2_{zzz}\left(\frac{5 (S^{zz})^2}{8}-\frac{3 z^2 S^{zz}}{8}+\frac{z^4}{16}\right)\ ,
\end{equation}  
which integrates to 
\begin{equation} 
F_2=C^2_{zzz}\left(\frac{5 (S^{zz})^3}{24}-\frac{3 z^2(S^{zz})^2}{16}+\frac{z^4 S^{zz}}{16} +\frac{z^6(729 z^2+162 z-11)}{1920}\right)\ .
\label{f2}
\end{equation} 
The integration constant $f_g$ of the $S^{zz}$ integration
($f_2=\frac{729z^2+162z-11}{1920(1+27 z)^2}$ in (\ref{f2})) can be fixed 
from the boundary behavior of ${\cal F}_g$. Since $z$  is a global parameter, we 
only need to know the holomorphic limit of $S^{zz}$ in terms of the flat 
coordinates $t_f\in\{T,t_c,\sigma\}$ near large radius, conifold and orbifold
point
\begin{equation} 
S^{zz}_f=\frac{2}{C_{zzz}} \partial_z {\cal F}^f_1=\frac{2}{C_{zzz}}
\partial_z\left(\frac{1}{2} \log\left(\frac{\partial z}{\partial
      t_f}\right)-\frac{1}{12}\log(z^7(1+ \frac{1}{27 z}))\right)\ 
\end{equation}
in order to evaluate ${\cal F}_g$ in the local coordinates in all patches.

The conditions on the local expansion  are similar as in the compact case in~\cite{HKQ},  
namely the gap condition at the  conifold, regularity at orbifold and the
constant map contribution at infinity. The difference is that in the
non-compact  case these conditions are sufficient to fix the kernel of 
(\ref{holomorphicanomaly}) completely. The argument is as follows.  
The maximal pole at the conifold is $\frac{1}{\Delta^{2g-2}(z)}$ and there is no
pole at the orbifold nor at infinity. Modularity implies that the possible
numerator of the ambiguity is a polynomial in the modular invariant
$z$. Since $F_g$ is finite at the orbifold at $\frac{1}{z}=0$ the denominator 
degree of $z$ cannot exceed $2g-2$, i.e. the ambiguity has to be of the form  
$\frac{p_{2g-2}(z)}{\Delta^{2g-2}}$. $2g-2$ of the $2g-1$ coefficients of 
$p_{2g-2}(z)$ follow from the gap condition
\begin{equation}  
\F_g= \frac{3^{g-1} B_{2g}}{2g (2g-2)t_c^{2g-2}}+{\cal O }(t_c^0),  
\label{gap}
\end{equation}  
here $t_c$ is the unique vanishing period at the conifold given in
(\ref{periodsconifold}). 
One additional condition follows from constant map contribution at infinity
\begin{equation} 
\F_g= \frac{3 B_{ 2 g-2} B_{2g}}{4 g (2 g-2) (2 g-2)!}+{\cal
  O}(Q)\ .   
\end{equation} 
With this boundary information the model is completely integrable. The integration
step can be further simplified.  As all $F_g$ are of the form $F_g=C_{zzz}^{2g-2}
P_g=C_{zzz}^{2g-2} \sum_{i=0}^{3g -3} (S^{zz})^i f_g^i(z)$, it is natural to 
rewrite (\ref{holomorphicanomaly}) for the $P_g$. To do this denote $\delta_z=\frac{1}{C_{zzz}} \partial_z$, so that e.g.
$\delta_z S^{zz}=(S^{zz})^2- z^2 (7+216 z)S^{zz}+\frac{z^4}{4}$, and
define the derivative $\delta$ on a weight $k$ function $g_k$ as 
$\delta g_k=(\delta_z+3 k z^2(1+36 z))g_k$. The weights are $[P_g]=6g-6$
and $[\delta P_g]=6g-3$ and (\ref{holomorphicanomaly}) reads
\begin{equation}  
\partial_{S^{zz}}  P_g = \frac{1}{2}\left(\left(\delta-\frac{\Gamma^z_{zz}}{C_{zzz}}\right) \delta P_{g-1}
  +\sum_{r=1}^{g-1} \delta P_{g-r}\delta P_{r}  \right)\ .
\label{closedanomaly} 
\end{equation}       
In this form the equation is most easily integrated to very high genus 
(up to genus $80$ in a few hours on a modern PC).

\subsection{Modular expressions for the $F_g$ on  $\K_{\P^2}$}\label{modexpP2}
The aim of this section is to relate the expression for $F_g$ obtained in  
the previous section to classical modular forms. Some results in this 
direction have been obtained in \cite{ABK} for a related  family of 
elliptic curves $\tilde \S(\tilde z)$ 
\begin{equation}
\sum_{i=1}^3 x_i^3+ {\tilde z}^{-\frac{1}{3}} \prod_{i=1}^3 x_i=0\ ,
\label{curve2}
\end{equation} 
which comes from the Landau-Ginzburg model, whose infrared limit  is the exact 
field theory $\C^3/\Z_3$ mentioned in the section \ref{orbifold}. 

In order to understand the relation between the curves let us calculate the
j-function of (\ref{curve2})
\begin{equation}
\tilde \jmath =\frac{(216 {\tilde z}-1)^3}{{\tilde z}(1+27\tilde{ z})^3}\ .
\label{j2}
\end{equation}
$\tilde \jmath$ is transformed into (\ref{j1}) when we identify
\begin{equation}
\tilde z=-\frac{1}{27}(1+27 z)\ 
\label{tzz}
\end{equation}
which exchanges the large radius point and the conifold point of $\tilde
\S(\tilde z)$ and $\S(z)$. Such reparametrization symmetries are ubiquitous 
in $N=2$ supersymmetric theories, e.g. in Seiberg-Witten  theory~\cite{Klemm:1995wp}, 
and the associated curves  $\S$ and $\tilde \S$  are called isogenous. 
It can be checked that periods of $\tilde \Sigma(\tilde z)$ fulfill the 
same Picard-Fuchs equation (\ref{pflocalp2}) as the ones of $\Sigma(z)$  
with the argument $z$ replaced by $\tilde z$. In fact the periods of the curves are 
related by a rescaling so that their modular parameter is rescaled 
by a factor $3$
\begin{equation}
\tau= 3 \tilde \tau  \ ,
\label{ttt} 
\end{equation}
as can be seen by comparing the $\tilde z(\tilde q)$ and $z(q)$ 
expansions that follow from (\ref{j2}) and (\ref{j1}).   

In \cite{ABK} quantities in the parameterization  of the curve (\ref{curve2}) 
have been related to $\theta$-constants that generate modular forms of
$\Gamma_0(3)$\footnote{~Because \cite{ABK} worked with (\ref{curve2}) all 
modular quantities below are understood to have the argument $\tilde \tau$.}
\begin{equation} 
a:=\theta^3\left[{1\over 6}\atop {1\over 6}\right], \quad
b:=\theta^3\left[{1\over 6}\atop {1 \over 2} \right], \quad
c:=\theta^3\left[{1\over 6}\atop {5\over 6}\right], \quad
d:=\theta^3\left[{1 \over 2} \atop {1\over 6}\right],
\end{equation}
which all have weight $3/2$ and satisfy with $\alpha=\exp\left(\frac{2 \pi i}{3}\right)$ 
the relations \cite{FK}
\begin{equation}
c=b-a, \qquad d= a+\alpha b, \qquad \eta^{12} = { i \over 3^{3/2}} a b c d \ .
\label{relations}
\end{equation}
Following the observation in \cite{ABK}  $\tilde \psi=-\frac{1}{\tilde
  z^{1/3}}=\alpha^2 \left(\frac{a-c-d}{d}\right)$ and
(\ref{relations}) we get 
\begin{equation}
\tilde z=-\frac{1}{3^3}\frac{d^4+\eta^{12}}{d^4}
\label{zoftau}
\end{equation}  
and 
\begin{equation}
\frac{\partial T}{\partial \tilde \psi}=-\alpha \sqrt{3} \frac{d}{\eta}\ ,  
\label{dTdpsi}
\end{equation}  
For this curve one finds the genus one amplitude 
\begin{equation}
F_1=-\log(\sqrt{\tilde \tau_2} \eta(\tilde \tau) \bar\eta(\bar {\tilde \tau}))+\frac{1}{24}\log\left(1+\frac{1}{27 \tilde
    z}\right)=-\frac{1}{2}\log\left({\tilde \tau}_2 \theta^{\frac{1}{2}}\left[{1
    \over 2} \atop {1\over 6}\right] \eta^\frac{1}{2} \bar \eta\right)  
\label{F1tau}
\end{equation}
Note that (\ref{F1tau}) can be transformed into (\ref{2F1tau}) by applying
(\ref{tzz}) and (\ref{ttt}). A small calculation using 
(\ref{F1tau},\ref{derF1}) and (\ref{dTdpsi}) gives the propagator in terms of
standard modular expressions 
\begin{equation}
S^{\tilde \psi \tilde \psi}=\left(\frac{\partial \tilde \psi}{\partial \tilde
    z}\right)^2\left(S^{zz}-\frac{{\tilde z}^2}{4}\right)=\frac{1}{12}
\left(\frac{\eta}{d}\right)^2 \hat E_2(\tilde \tau)\ .
\label{smodular} 
\end{equation}
This and  (\ref{zoftau}) allows to rewrite all $F_g$ in terms of theta
functions and $\hat E_2$. With $F_g=X^{g-1} \hat P_g$, where $X=\frac{d^2}{2^9
3^{6} \eta^{18}}$ is a weight $-3$ form, we get  e.g. 
\begin{equation}
\hat P_2=5 \hat E_2^3+\frac{\alpha}{\eta^2}\left(\frac{d^4+27
    \eta^{12}}{d}\right)^\frac{2}{3} \hat E_2^2-\frac{\alpha^2}{3 \eta^4}\left(\frac{d^4+27
    \eta^{12}}{d}\right)^\frac{4}{3} \hat E_2-\frac{(d^4-27 \eta^{12})(d^4+33
    \eta^{12})}{15 d^2 \eta^2}\ .
\end{equation}
Since $\hat E_2,d,\eta$ close under derivatives  $d_{\tilde \tau} d=\frac{E_2
  d}{8}+\frac{d^3}{108 \eta^2 (-\tilde z)^\frac{2}{3}}$ ($d_{\tilde \tau}
\tilde z=- 3^3\frac{\eta^{10}}{d^2} (-\tilde z)^\frac{4}{3}$), it is obviously 
possible to set up the direct integration in terms of the modular expression. 
We leave this to the reader.

\subsection{The higher genus results for $\K_{\P^2}$}

At the large radius point we recorded some Gopakumar-Vafa invariants in 
appendix \ref{GVInvariants}. The results agree with the literature
as far as they are known. Both w.r.t. to the  genus as well as to the 
degree the method outlined here is the most effective one to get these 
generating functions. An excellent check on this data is provided already by 
the formulas $n^{g(d)}_d=(-1)^{\frac{d(d+3)}{2}}\frac{(d+1)(d+2)}{2}$ and 
$n^{g(d)-1}_d=-(-1)^{\frac{d(d+3)}{2}}\left(d\atop 2\right)(d^2+d-3)$
for the highest genus  $g(d)=\frac{(d-1)(d-2)}{2}$ and the next to highest genus 
BPS invariant in each degree $d$, which were derived in~\cite{Katz:1999xq}. 
In fact we checked that the spaces in~\cite{Katz:1999xq}, which model 
the moduli space of the $D_2$-$D_0$ brane system with $D_2$ brane 
charge $d$ are smooth for $D_2$ branes wrapping  holomorphic curves 
of genus $g(d)-\delta$ with up to $\delta=d-1$ nodes. As a consequence 
the formula (4.15) of~\cite{Katz:1999xq} applies for $n^{g(d)-\delta}_d$, 
with $e({\cal C}^{(p)})= e(\mathbb{P}^{(d (d + 3)/2 - p})e({\rm Hilb}^p 
\mathbb{P}^2)$ for $\delta=0,\dots,d-1$, yielding $120$ non-trivial 
checks for the BPS numbers in appendix \ref{GVInvariants}. We also expect 
that the relatively simple recursive nature of the procedure 
described here will allow to study high genus asymptotics of 
BPS states.   

The ${\cal F}^c_g$  near the conifold are expected to correspond to a perturbation of the  
$c=1$ string at selfdual radius, which has been established as a dual 
description of the topological string at the conifold~\cite{Ghoshal:1995wm},
but the details of the identification of the perturbation parameters are not completely 
clarified~\cite{Dijkgraaf:2003xk}. The most notable structure is the gap 
in the ${\cal F}_g^{c}$ expansion at higher genus. We display a few low genus ${\cal F}^c_g$  
\begin{equation} 
\begin{array}{rl}
{\cal F}^{c}_2&=
\frac{1}{80\,t_c^2} - \frac{1}{51840} - 
\frac{t_c}{19440} + \frac{3187 t_c^2}{377913600} - \frac{239\,t_c^3}{255091680} + {\cal O}(t_c^4)\\
{\cal F}^{c}_3&=
\frac{1}{112\,t_c^4} - \frac{1}{117573120} - \frac{t_c}{1469664} + \frac{23855\,t_c^2}{179992689408} - \frac{557\,t_c^3}{24794911296} + {\cal O}(t_c^4)\\
{\cal F}^{c}_4&=\frac{3}{160\,t_c^6} - \frac{1}{63489484800} - \frac{7\,t_c}{377913600} + \frac{6830569\,t_c^2}{1190155742208000} - 
  \frac{1561279\,t_c^3}{1205032688985600} + {\cal O}(t_c^4)\\
{\cal F}^{c}_5&=\frac{27}{352\,t_c^8} - \frac{1}{16761223987200} - \frac{809\,t_c}{942818849280} + \frac{118418785\,t_c^2}{326612060022657024} - 
  \frac{113975899\,t_c^3}{1002105184160424960} + {\cal O}(t_c^4) \\
{\cal F}^{c}_6&=\frac{18657}{36400\,t_c^{10}} - \frac{691}{1853204730144768000} - \frac{1276277\,t_c}{21059144660736000} + 
  \frac{279842720162009\,t_c^2}{9052836032762704465920000}+{\cal O}(t_c^3) \\
{\cal F}^{c}_7&=\frac{81}{16\,t_c^{12}} - \frac{691}{200146110855634944000} - \frac{7943\,t_c}{1309171316428800} + 
  \frac{27776712091\,t_c^2}{7792369912031464488960}  +  {\cal O}(t_c^3) \\
{\cal F}^{c}_8&=\frac{2636793}{38080\,t_c^{14}} - \frac{3617}{81659613229099057152000} - \frac{25034924437\,t_c}{30622354960912146432000} +  {\cal O}(t_c^2)  
\end{array}
\label{closedstringconfold}
\end{equation}

If we denote as in~\cite{ABK} the generating function
\begin{equation}
{\cal F}_g^{orb}=\frac{1}{(3 k)!} N_{g,k} \sigma^{3 k}\ ,
\end{equation} 
we can read of the orbifold Gromov-Witten invariants, see \cite{ABK}\cite{BC},
from our results, as in the table below.\footnote{~It corrects some misprints in \cite{ABK}\cite{BC}.}
\begin{table}[h]
\centering
{\footnotesize{
\begin{tabular}[h]{c|ccccc} 
\toprule
$g\backslash d$&0&1&2&3&4 \\
\midrule
0&$ $&$\frac{1}{3}$&$ \frac{-1}{3^3}$&$ \frac{1}{3^2}$&$ \frac{-1093}{3^6}$\\[2 mm]
1&$ $&$ 0$&$ \frac{1}{3^5}$&$ \frac{-14}{3^5}$&$ \frac{13007}{3^8}$\\ [2 mm]
2&$\frac{1}{2^7 3^3 5}$&$ \frac{1}{2^4 3^5 5}$&$ \frac{-13}{2^4 3^6}$&$
\frac{20693}{2^4 3^8 5}$&$ \frac{-12803923}{2^4 3^10 5}$\\ [2 mm]
3&$\frac{-1}{2^9 3^5 5\cdot 7}$&$ \frac{-31}{2^5 3^7 5\cdot 7}$&$ \frac{11569}{2^5 3^9 5\cdot 7}$&$ \frac{-2429003}{2^5 3^10 5\cdot 7}$&$ \frac{871749323}{2^5 3^11 5\cdot 7}$\\ [2 mm]
4&$\frac{-311}{2^11 3^8 5^2 7}$&$ \frac{313}{2^7 3^9 5^2}$&$ \frac{-1889}{2^83^9}$&$ \frac{115647179}{2^6 3^13 5^2}$&$ \frac{-29321809247}{2^8 3^12 5^2}$\\[2 mm] 
5&$\frac{24559}{2^14 3^9 5^2 7\cdot 11}$&$ \frac{-519961}{2^9 3^11 5^2 7\cdot11}$&$ \frac{196898123}{2^9 3^12 5^2 7\cdot 11}$&$ \frac{-339157983781}{2^93^14 5^2 7\cdot 11}$&$ \frac{78658947782147}{2^9 3^16 5^2 7}$\\[2 mm] 
6&$\frac{-49922143}{2^14 3^11 5^3 7^2 11\cdot 13}$&$ \frac{14609730607}{2^12
  3^13 5^3 7^2 11\cdot 13}$&$ \frac{-258703053013}{2^10 3^15 5\cdot 7^2 11\cdot
  13}$&$ \frac{2453678654644313}{2^12 3^14 5^3 7^2 11\cdot 13}$&$ 
\frac{-4001577419369601803}{2^11 3^18 5^3 7^2 11\cdot 13}$\\ [2 mm]
7&$\frac{1341390269}{2^16 3^13 5^3 7^2 11\cdot 13}$&$ 
  \frac{-1122101011}{2^13 3^14 5^3 7\cdot 11}$&$ 
  \frac{2196793414201}{2^11 3^17 5^3 7\cdot 11}$&$ 
  \frac{-2127526097369539}{2^13 3^18 5^2 7\cdot 11}$&$ 
  \frac{26373375124439869913}{2^12 3^20 5^3 7\cdot 11}$\\ [2 mm]
8&$\frac{-1701146456533}{2^19 3^15 5^3 7^2 11\cdot 13\cdot 17}$&$ 
  \frac{1424424798274897}{2^15 3^17 5^4 7^2 11\cdot 13\cdot 17}$&$ 
  \frac{-80699319730594681}{2^15 3^19 5^3 7^2 11\cdot 17}$&$ 
  \frac{3471527490671857976969}{2^16 3^20 5^3 7^2 11\cdot 13\cdot 17}$&$ \frac{-114258620434929543630324227}{2^16 3^22 5^4 7^2 11\cdot 13\cdot 17}$\\
\bottomrule
\end{tabular}}}
\caption{Low genus orbifold Gromov-Witten invariants  $N_{g,d}$}
\end{table} 
Some of the results beyond $g=0$ have been confirmed in \cite{BC}.

\section{$\K_{\P^1 \times \P^1}=\cO(-2,-2)\rightarrow\P^1\times\P^1$}
\label{localF0} 
We are considering the non-compact Calabi-Yau geometry $\cO(-2,-2)\rightarrow\P^1\times\P^1$, i.e.~the canonical line bundle over the Hirzebruch surface $\HirzeF_0=\P^1\times\P^1$. This local model can be obtained from the compact elliptic fibration over $\HirzeF_0$ with fiber $X_6(1,2,3)$. The three complexified K\"ahler volumes have the corresponding Mori cone generators $(-6;3,2,1,0,0,0,0)$, $(0;0,0,-2,1,0,1,0)$, $(0;0,0,-2,0,1,0,1)$. Roughly, in the local limit the volume of the elliptic fiber is send to infinity. The B-model mirror description of the local geometry is encoded in a Riemann surface with a meromorphic differential as pointed out before.\\
According to \cite{HKTY} and using the above mentioned charge vectors, one can derive a Picard-Fuchs system governing the periods of the global mirror geometry. They are given by
\begin{equation}
\begin{split}
&\cD_1=\Theta_1(\Theta_1-2\Theta_2-2\Theta_3)-18z_1(1+6\Theta_1)(5+6\Theta_1)\\
&\cD_2=\Theta_2^2+z_2(1-\Theta_1+2\Theta_2+2\Theta_3)(\Theta_1-2\Theta_2-2\Theta_3)\\
&\cD_3=\Theta_3^2+z_3(1-\Theta_1+2\Theta_2+2\Theta_3)(\Theta_1-2\Theta_2-2\Theta_3),
\end{split}
\end{equation}
where we denote the logarithmic derivative by $\Theta_i=z_i\frac{\d}{\d z_i}$. $z_1$ is the complex structure parameter dual to the K\"ahler parameter of the elliptic fiber $t_{\rm F}$. The local limit is obtained by sending this parameter to zero, $z_1\rightarrow0$.\\
Now let us turn to the non-compact geometry. The toric data of local $\HirzeF_0$ is summarized in the following matrix, $V$ denoting the vectors which span the fan and $Q$ denoting the charge vectors.
\begin{equation}
(V|Q)=\left(\begin{array}{ccc|cc} 0&0&1&-2&-2\\1&0&1&1&0\\ 0&-1&1&0&1\\ -1&0&1&1&0 \\ 0&1&1&0&1 \end{array}\right)
\end{equation}
From there we conclude the following quantities as was explained in section \ref{bmodel}. $C^{(0)}_{ijk}$ denote the classical triple intersection numbers. They, as well as $\int_M c_2J_i$, were computed using toric geometry.
\begin{equation} \label{topdata}
\begin{array}{ll}
a) & Q^1=(-2,1,0,1,0),\; Q^2=(-2,0,1,0,1) \\
b) & Z=\{x_1=x_3=0\}\cup\{x_2=x_4=0\} \\
c) & M=(\C^5[x_0,\cdots,x_4]\setminus Z)/(\C^*)^2 \\
d) & H(x,y)=y^2-x^3-(1-4z_1-4z_2)x^2-16z_1z_2x \\
e) & \cD_1=\Theta_1^2-2z_1(\Theta_1+\Theta_2)(1+2\Theta_1+2\Theta_2)\\
   & \cD_2=\Theta_2^2-2z_2(\Theta_1+\Theta_2)(1+2\Theta_1+2\Theta_2)\\
   & \Delta=1-8(z_1+z_2)+16(z_1-z_2)^2 \\
f) & C^{(0)}_{111}=\frac{1}{4},\; C^{(0)}_{112}=-\frac{1}{4},\; C^{(0)}_{122}=-\frac{1}{4},\; C^{(0)}_{222}=\frac{1}{4}\\
g) & \int\limits_M c_2J_1=\int\limits_M c_2J_2=-1.
\end{array}
\end{equation}
$H(x,y)=0$ defines a family of elliptic curves $\Sigma(z_1,z_2)$ whose $j$-function is given by
\begin{equation}
j(z_1,z_2)=\frac{((1-4z_1-4z_2)^2-48z_1z_2)^3}{z_1^2z_2^2(1-8(z_1+z_2)+16(z_1-z_2)^2)}.
\end{equation}

\subsection{Review of the moduli space $\M$}
The moduli space, $\M$, of the local Calabi-Yau $\cO(-2,-2) \rightarrow \P^1 \times \P^1$ is spanned by two K\"ahler moduli controlling the sizes of the two $\P^1$'s. The B-model mirror description of this geometry can be expressed through a Riemann surface together with a meromorphic differential. The meromorphic differential is the reduction of the holomorphic three-form of the mirror geometry to a one-form living on a Riemann surface as described in section \ref{bmodel}. In our particular case we get a genus one Riemann surface with two non-trivial cycles. Apart from these the meromorphic differential has a residue arising from integration over a certain trivial cycle. Together these periods parameterize the two complex structure moduli which are mirror to the two K\"ahler moduli of the original model. The period integrals satisfy two linear differential equations of order two, given by the Picard-Fuchs operators. It is well known that these periods can at worse have logarithmic singularities. The singular locus in the moduli space can be obtained by calculating the discriminant of the Picard-Fuchs system (\ref{topdata}). This yields
\begin{equation} \label{singloc}
z_1 z_2 \left(1-8 (z_1 + z_2) + 16 (z_1-z_2)^2\right)=:z_1 z_2 \Delta = 0.
\end{equation}
One sees that the singular locus splits into three irreducible components given by the divisors $z_1=0$, $z_2=0$ and $\Delta=0$. The moduli $z_1,z_2$ are compactified to $\P^2$.\\
At the large complex structure point $L_1 \cap L_2$, two of the periods, $t_1=\log(z_1) + \cO(z)$ and $t_2=\log(z_2) + \cO(z)$, give the classical large K\"ahler volumes of the two $\P^1$. As $C$ touches $L_1$ at $z_2=\frac{1}{4}$, $L_2$ at $z_1=\frac{1}{4}$ and $I$ at $u=\frac{z_1}{z_1+z_2}=\frac{1}{2}$ and all intersections are with contact order two, the Picard-Fuchs system cannot be solved around these points in moduli space. Therefore, the moduli space has to be blown up around these points so that all divisors have normal crossings. This is done by introducing two new divisors at each of these points which is depicted in figure \ref{rms}.
\begin{figure} 
\center
\begin{picture}(0,0)%
\includegraphics{F0ModuliSpace.pstex}%
\end{picture}%
\setlength{\unitlength}{1865sp}%
\begingroup\makeatletter\ifx\SetFigFont\undefined%
\gdef\SetFigFont#1#2#3#4#5{%
  \reset@font\fontsize{#1}{#2pt}%
  \fontfamily{#3}\fontseries{#4}\fontshape{#5}%
  \selectfont}%
\fi\endgroup%
\begin{picture}(8832,7234)(-764,-7238)
\put(4006,-6541){\makebox(0,0)[lb]{\smash{{\SetFigFont{9}{10.8}{\rmdefault}{\mddefault}{\updefault}$E_1$}}}}
\put(901,-6901){\makebox(0,0)[lb]{\smash{{\SetFigFont{9}{10.8}{\rmdefault}{\mddefault}{\updefault}$L_1=\{z_1=0\}$}}}}
\put(-764,-6136){\makebox(0,0)[lb]{\smash{{\SetFigFont{9}{10.8}{\rmdefault}{\mddefault}{\updefault}$L_2=\{z_2=0\}$}}}}
\put(6391,-4696){\makebox(0,0)[lb]{\smash{{\SetFigFont{9}{10.8}{\rmdefault}{\mddefault}{\updefault}$I=\{\frac{1}{z_1+z_2}=0\}$}}}}
\put(3241,-7171){\makebox(0,0)[lb]{\smash{{\SetFigFont{9}{10.8}{\rmdefault}{\mddefault}{\updefault}$F_1$}}}}
\put(451,-4471){\makebox(0,0)[lb]{\smash{{\SetFigFont{9}{10.8}{\rmdefault}{\mddefault}{\updefault}$F_2$}}}}
\put(5266,-2491){\makebox(0,0)[lb]{\smash{{\SetFigFont{9}{10.8}{\rmdefault}{\mddefault}{\updefault}$F$}}}}
\put(5266,-3346){\makebox(0,0)[lb]{\smash{{\SetFigFont{9}{10.8}{\rmdefault}{\mddefault}{\updefault}$E$}}}}
\put(2836,-3526){\makebox(0,0)[lb]{\smash{{\SetFigFont{9}{10.8}{\rmdefault}{\mddefault}{\updefault}$C$}}}}
\put(1126,-5056){\makebox(0,0)[lb]{\smash{{\SetFigFont{9}{10.8}{\rmdefault}{\mddefault}{\updefault}$E_2$}}}}
\end{picture}%
\caption{Resolved Moduli Space of $\HirzeF_0$}
\label{rms}
\end{figure}
More details about this moduli space can be found in \cite{Aganagic:2002wv}. For us the most relevant points are $I \cap F$ which is a $\Z_2$ orbifold point admitting a matrix model expansion, and the conifold locus $C$, relevant for fixing the holomorphic ambiguity of the free energy functions.

\subsection{Solving the topological string on local $\HirzeF_0$ at large radius}

By the method of Frobenius one can calculate the periods eliminated by the Picard-Fuchs system. As the charge vectors are chosen such that they span the Mori cone, the periods are calculated at the large radius point of the moduli space $\M(M)$. It is well known that the regular solution for this local model is simply $\omega_0(\underline{z},0)=1$. Therefore the mirror map is equal to the single logarithmic solution and given by
\begin{equation}
\begin{array}{l}
2\pi i T_1(z_1,z_2)=\log z_1+2(z_1+z_2)+3(z_1^2+4z_1z_2+z_2^2)+\frac{20}{3}(z_1^3+9z_1^2z_2+9z_1z_2^2+z_2^3)+\cO(z^4)\\
2\pi i T_2(z_1,z_2)=\log z_2+2(z_1+z_2)+3(z_1^2+4z_1z_2+z_2^2)+\frac{20}{3}(z_1^3+9z_1^2z_2+9z_1z_2^2+z_2^3)+\cO(z^4).\\
\end{array}
\end{equation}
By inverting the above series we arrive at ($Q_i=e^{2\pi i T_i}$)
\begin{equation}
\begin{array}{l}
z_1(Q_1,Q_2)=Q_1-2(Q_1^2+Q_1Q_2)+3(Q_1^3+Q_1Q_2^2)-4(Q_1^4+Q_1^3Q_2+Q_1^2Q_2^2+Q_1Q_2^3)+\cO(Q^5)\\
z_2(Q_1,Q_2)=Q_2-2(Q_1Q_2+Q_2^2)+3(Q_1^2Q_2+Q_2^3)-4(Q_1^3Q_2+Q_1^2Q_2^2+Q_1Q_2^3+Q_2^4)+\cO(Q^5).
\end{array}
\end{equation}
We observe that the following combination does not receive any instanton corrections which can be easily derived from the Picard-Fuchs system
\begin{equation}
\frac{z_1}{z_2}=\frac{Q_1}{Q_2}=e^{2\pi i (T_1-T_2)}=:Q^x_1,
\end{equation}
or in other words, the mirror map can be brought in trigonal form by means of the coordinate choice, $x_1=\frac{z_1}{z_2}$ and $x_2=z_2$, as well as $Q^x_2=Q_2$. We have
\begin{equation}
\begin{array}{l}
x_1(Q^x_1,Q^x_2)=Q^x_1,\\
x_2(Q^x_1,Q^x_2)=Q^x_2-2{Q^x_2}^2+Q^x_1{Q^x_2}^2+3{Q^x_2}^3+\cO(Q^4).
\end{array}
\end{equation}
The next step is to determine the Yukawa couplings. Four independent combinations are
\begin{eqnarray}
C_{111}=\frac{(1-4z_2)^2-16z_1(1+z_1)}{4z_1^3\Delta},\quad 
C_{112}=\frac{16z_1^2-(1-4z_1)^2}{4z_1^2z_2\Delta},\nonumber\\
C_{122}=\frac{16z_2^2-(1-4z_2)^2}{4z_1z_2^2\Delta},\quad 
C_{222}=\frac{(1-4z_1)^2-16z_1(1+z_2)}{4z_2^3\Delta}.
\end{eqnarray}
The numerator is fixed by the help of the known classical triple intersection numbers as well as the genus zero GV invariants, whereas the denominator is fixed by the Picard-Fuchs system. Note, that the Yukawa couplings are of the well-known structure, i.e. a rational function in the $z_i$'s multiplied by the inverse of the discriminant. Here we note, that in local models the choice of the classical data is crucial for the success of direct integration. This is due to the fact, that one can obtain the right GV invariants for different choices of $C^{(0)}$ and $\int c_2J$. However, if one does not use consistent data, higher genus calculations become wrong or even impossible. In contrast, the dependence on some Euler number drops out completely, as it does not effect the GV invariants. In this work we simply set $\chi$ to zero.\\
Using the ansatz (\ref{1F1}) for the free energy function of genus one and the classical data $\int c_2J_i$ as well as the known genus one GV invariants we are able to fix the holomorphic ambiguity at genus one, $f_1$. The result as well as the expansion at large radius in the holomorphic limit $\overline{T}\rightarrow0$ reads as follows
\begin{equation}
\begin{split}
F_1&=\log\left(\Delta^{-\frac{1}{12}}(z_1z_2)^{-\frac{13}{24}}(\det (G_{i\bar\jmath}))^{-\frac{1}{2}}\right), \\
\F_1 (T_1,T_2)&=-\frac{1}{24}\log(Q_1Q_2)-\frac{1}{6}(Q_1+Q_2)-\frac{1}{12}(Q_1^2+4Q_1 Q_2+Q_2^2)+\cO(Q^3).
\end{split}
\end{equation}
In order to perform the method of direct integration, we have to calculate the propagator and express all quantities which carry non-holomorphic information through our propagators. As a first step the holomorphic ambiguity, $\tilde{f}$, in (\ref{PropEq}) can be fixed by the choice
\begin{eqnarray}
\tilde{f}^1_{11}=-\frac{1}{z_1},\;\tilde{f}^1_{12}=-\frac{1}{4z_2},\;\tilde{f}^1_{22}=0,\nonumber\\
\tilde{f}^2_{11}=0,\;\tilde{f}^2_{12}=-\frac{1}{4z_1},\;\tilde{f}^2_{22}=-\frac{1}{z_2},
\end{eqnarray}
where all other combinations follow by symmetry. We note that the propagator has only one independent component for we can write
\begin{equation}
S^{ij}=\begin{pmatrix}
S(z_1,z_2) & \displaystyle\frac{z_2}{z_1}\,S(z_1,z_2) \\
\displaystyle\frac{z_2}{z_1}\,S(z_1,z_2) & \displaystyle\frac{z_2^2}{z_1^2}\,S(z_1,z_2)
       \end{pmatrix}
\end{equation}
where $S(z_1,z_2)=\frac{1}{2}z_1^2-2z_1^3-2z_1^2z_2-8z_1^3z_2-32z_1^4z_2+\cO(z^6)$. This is due to the fact, that the mirror geometry is solely determined by the elliptic curve $\Sigma(z_1,z_2)$, which has only one relevant elliptic parameter $\tau$. The dependence on a second parameter is due to a non-vanishing residue of the meromorphic differential on $\Sigma(z_1,z_2)$.\\
Often it is convenient and also more natural to perform the calculations in the coordinates $x_1,x_2$, in which some Christoffel symbols are rational
$$ \Gamma^1_{11}=\frac{1}{x_1},\quad \Gamma^1_{12}=0,\quad \Gamma^1_{22}=0. $$
Noting, that from the tensorial transformation law of the propagator and the relation (\ref{PropEq}) the ambiguity of the propagator $\tilde{f}$ has to transform as $\tilde{f}^i_{jk}(x)=\frac{\d x_i}{\d z_l}(\frac{\d^2 z_l}{\d x_j\d x_k})+\frac{\d x_i}{\d z_l}\frac{\d z_m}{\d x_j}\frac{\d z_n}{\d x_k}\tilde{f}^l_{mn}(x(z))$. We obtain
\begin{equation}
\tilde{f}^1_{11}=-\frac{1}{x_1},\;\tilde{f}^2_{12}=-\frac{1}{4x_1},\;\tilde{f}^2_{22}=-\frac{3}{2x_2},
\end{equation}
where all other combinations are either 0 or follow by symmetry. As $\Gamma^1_{ij}=-\tilde{f}^1_{ij}$ we observe that the propagator takes the following simple form $S^{11}=S^{12}=S^{21}=0$ and $S^{22}=\frac{x_2^2}{2}-2x_2^3-2x_1x_2^3+\cO(x^5)$.\\
In addition, we fix the holomorphic ambiguity of the covariant derivative of $S^{ij}$, (\ref{DS}), and obtain
\begin{equation}
\begin{array}{l}
f_1^{11}=-\frac{1}{8}z_1(1+4z_1-4z_2),\;f_1^{12}=-\frac{1}{8}z_2(1+4z_1-4z_2),\;f_1^{22}=-\frac{z_2^2}{8z_1}(1+4z_1-4z_2),\\
f_2^{11}=-\frac{z_1^2}{8z_2}(1+4z_2-4z_1),\;f_2^{12}=-\frac{1}{8}z_1(1+4z_2-4z_1),\;f_2^{22}=-\frac{1}{8}z_2(1+4z_2-4z_1),
\end{array}
\end{equation}
where all other combinations follow by symmetry. Further we can express the covariant derivative of $F_1$ through the generator $S$ (\ref{F1prop}) by
\begin{equation}\label{F1propF0}
D_iF_1=\frac{1}{2}C_{ijk}S^{jk}-\frac{1}{12}\Delta^{-1}\d_i\Delta+\frac{7}{24z_i}.
\end{equation}
Note, that in contrast to an one parameter model like in section \ref{localp2} the holomorphic ambiguity $A_i=\d_i(\tilde{a}_j\log\Delta_j+\tilde{b}_j\log z_j)$ in (\ref{F1propF0}) cannot be set to zero. More generally, in the local models we are considering here the geometry of the B-model is encoded in a Riemann surface of genus one whose moduli space admits only one quasimodular form of weight 2, namely the second Eisenstein series. Therefore and from the discussions in the case of local $\P^2$ in the previous section we expect there to be a coordinate system in which the propagator is proportional to the second Eisenstein series. The relevant coordinate system is given by the $x$-coordinates in which it is allowed to set all but one component of the propagator to zero and subsequently one can use (\ref{F1prop}) and (\ref{1F1}) to solve for this non-zero component. Now, in the multi-parameter case this gives, for each direction of the derivative of $F_1$ w.r.t.~$z_i$, $h^{2,1}$ equations on $\tilde{a}_j,\tilde{b}_j$. In this and the following example, we are lucky as these constraints fix the parameters completely. In addition one arrives at a series expansion for the non-vanishing component of $S^{ij}$. This can be used to fix all ambiguities in the model as rational functions of the $z_i$ with poles only at the singular divisors of the Picard-Fuchs system.\\

Now, all input to perform direct integration is provided and applying this method we are able to determine $F_g$ for genus $g$ up to four. Using that local $\HirzeF_0$ has a discriminant with $\deg\Delta=2$ and we can further reduce the number of coefficients in $A_g$ due to symmetry in $z_1$ and $z_2$, one can easily calculate, that at genus $g$ there are $(2g-1)^2$ unknowns in the holomorphic ambiguity. Therefore genus four corresponds to fixing 49 coefficients in the holomorphic ambiguity $f_g=\frac{A_g}{\Delta^{2g-2}}$. They are determined by the gap condition at the conifold locus and the known constant map contributions. We will further comment on this in the next section.\\

Let's present at least the genus two results. The free energy is given by
\begin{equation}
\begin{split}
F_2=&\frac{5}{24z_1^6\Delta^2}S^3+\frac{-13+48z_1^2+z_1(40-96z_2)+40z_2+48z_2^2}{48z_1^4\Delta^2}S^2+\\ &\frac{384z_1^3+z_1^2(80-384z_2)+(1-4z_2)^2(17+24z_2)-16z_1(7-46z_2+24z_2^2)}{144z_1^2\Delta^2}S+f_2,
\end{split}
\end{equation}
where the ambiguity $f_2=\frac{A_2}{\Delta^2}$ is fixed by the following choice
\begin{equation}
\begin{split}
A_2=&-\frac{1}{1440}(25-258z_1+696z_1^2+416z_1^3-2688z_1^4-258z_2+2768z_1z_2-6560z_1^2z_2-1536z_1^3z_2\\
&+696z_2^2-6560z_1z_2^2+8448z_1^2z_2^2+416z_2^3-1536z_1z_2^3-2688z_2^4).
\end{split}
\end{equation}
The solution around the conifold is described in the next section. The GV
invariants can be found in the appendix \ref{GVInvariants}. They are in accord
with~\cite{Aganagic:2002qg} 
as far as they have been computed.

\subsection{Solving the topological string on local $\HirzeF_0$ at the conifold locus}
\begin{figure}
\center
\begin{picture}(0,0)%
\includegraphics{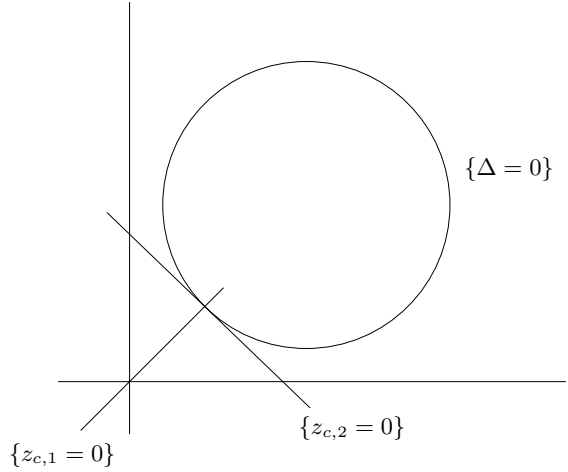}%
\end{picture}%
\setlength{\unitlength}{2072sp}%
\begingroup\makeatletter\ifx\SetFigFont\undefined%
\gdef\SetFigFont#1#2#3#4#5{%
  \reset@font\fontsize{#1}{#2pt}%
  \fontfamily{#3}\fontseries{#4}\fontshape{#5}%
  \selectfont}%
\fi\endgroup%
\begin{picture}(6672,5578)(2206,-6257)
\put(7651,-2761){\makebox(0,0)[lb]{\smash{{\SetFigFont{9}{10.8}{\rmdefault}{\mddefault}{\updefault}$\{\Delta = 0\}$}}}}
\put(5671,-5866){\makebox(0,0)[lb]{\smash{{\SetFigFont{9}{10.8}{\rmdefault}{\mddefault}{\updefault}$\{z_{c,2}=0\}$}}}}
\put(2206,-6181){\makebox(0,0)[lb]{\smash{{\SetFigFont{9}{10.8}{\rmdefault}{\mddefault}{\updefault}$\{z_{c,1}=0\}$}}}}
\end{picture}%
\caption{Conifold coordinates}
\label{conifig}
\end{figure}

Our next task is to solve the Picard-Fuchs equations around the conifold locus. In order to do that we choose some convenient point on the locus and define variables which are good coordinates around this point. In our case we choose the point to be $z_1=\frac{1}{16}, z_2=\frac{1}{16}$. As one can easily check inserting these numbers into the discriminant yields zero. To find the right variables we have to be careful as their gradients at the relevant point must not be colinear. The following choice will do the job
\begin{equation}\label{conicoord}
\cz{1}=1-\frac{z_1}{z_2},\quad \cz{2}=1-\frac{z_2}{\frac{1}{8}-z_1}.
\end{equation}
We transform the Picard-Fuchs system to the above coordinates and find the following polynomial solutions
\begin{eqnarray}
\omega^c_0 & = & 1, \nonumber \\
\omega^c_1 & = & -\log(1-\cz{1}),\nonumber \\
\omega^c_2 & = & \cz{2} + \frac{1}{16}(2\cz{1}^2+8\cz{1}\cz{2}+13\cz{2}^2)+\cO(z_c^3).
\end{eqnarray}
As mirror coordinates we take $\ct{1}:=\omega^c_1$ and $\ct{2}:=\omega^c_2$. Inverting these series gives the following mirror map
\begin{eqnarray}
\cz{1}(\ct{1},\ct{2}) &=& 1-e^{-\ct{1}},\nonumber \\
\cz{2}(\ct{1},\ct{2}) &=& \ct{2}-\frac{1}{16}(\ct{1}^2+8\ct{1}\ct{2}+13\ct{2}^2)+\cO(t_c^5).
\end{eqnarray}
The divisor $\{\cz{1}=0\}$ is normal to the conifold locus at $(z_1,z_2)=\left(\frac{1}{16},\frac{1}{16}\right)=p_{\rm con}$ whereas $\{\cz{2}=0\}$ is tangential (see figure \ref{conifig}). Therefore $\cz{1}$ parameterizes the tangential direction to the conifold locus at $p_{\rm con}$ in moduli space and $\cz{2}$ the normal one. Hence we expect the flat mirror coordinate $\ct{2}$ to be controlling the size of the shrinking cycle at $p_{\rm con}$, thus $\ct{2}$ should appear in inverse powers in the expansion of the free energies.\\

Transforming the Yukawa couplings, the Christoffel symbols and the holomorphic ambiguities $\tilde{f}$ to the conifold coordinates we obtain the propagator around this locus. In the choice of our coordinates (\ref{conicoord}) the propagator takes the following simple form $ S^{11}=S^{12}=S^{21}=0$ and 
$$ S^{22}=\half\ct{2}+\frac{1}{1536}(24\ct{1}^2\ct{2}+\ct{2}^3) +\cO(t_c^4). $$
Assuming the gap condition holds, we are able to fix all but one coefficients of the holomorphic ambiguity. Expanding the free energies at the large radius point in moduli space the constant map contribution fixes the last unknown, i.e.~we observe that the gap condition yields at genus two 8 out of 9 unknowns, at genus three 24 out of 25 unknowns, etc. Our results up to genus four are given below (rescaling: $\ct{2}\rightarrow2\ct{2}$)
\begin{equation}
\begin{split}
\F^c_2&=-\frac{1}{240\ct{2}^2}-\frac{1}{1152}+\frac{53\ct{2}}{122880}+\frac{\ct{1}^2}{61440}-\frac{2221\ct{2}^2}{14745600}+\cO(t_c^3)\\
\F^c_3&=\frac{1}{1008\ct{2}^4}+\frac{23}{5806080}+\frac{407\ct{2}}{198180864}-\frac{\ct{1}^2}{3096576}-\frac{258485\ct{2}^2}{49941577728}+\cO(t_c^3)\\
\F^c_4&=-\frac{1}{1440\ct{2}^6}-\frac{19}{278691840}+\frac{114773\ct{2}}{362387865600}+\cO(t_c^2).
\end{split}
\end{equation}

\subsection{Solving the topological string on local $\HirzeF_0$ at the orbifold point}
As we have noted already there exists an orbifold point in the moduli space $\M$ at which we can compare our results with the known matrix model expansions.\\
At this point we expand the periods in the local variables
\begin{equation}\label{orbicoord}
\oz{1}=1-\frac{z_1}{z_2},\quad \oz{2}=\frac{1}{\sqrt{z_2}\left(1-\frac{z_1}{z_2}\right)}.
\end{equation}
Transforming the Picard-Fuchs system to these coordinates and solving it, we obtain the following set of periods
\begin{eqnarray}
\omega^o_0 & = & 1, \nonumber \\
\omega^o_1 & = & -\log(1-\oz{1}), \nonumber \\
\omega^o_2 & = & \oz{1}\oz{2}+\frac{1}{4}\oz{1}^2\oz{2}+\frac{9}{64} \oz{1}^3 \oz{2} + \cO(z_o^5), \nonumber \\
F^{(0)}_{\omega^o_2}&=&\omega^o_2\log(\oz{1})+\frac{1}{2}\oz{1}^2\oz{2}+\frac{21}{64}\oz{1}^3\oz{2}+\cO({z_o}^5).
\end{eqnarray}
We define the mirror map to be given by the first two periods
\begin{equation}
  \ot{1}:=\omega^o_1,\quad \ot{2}:=\omega^o_2,
\end{equation}
and will express the B-model correlators in terms of these coordinates. In order to invert the mirror map and find the function $z_o(t_o)$, we have to consider the two series $\ott{1}=\ot{1}=\oz{1}+1+\cO({z_o}^2)$ and $\ott{2}=\frac{\ot{2}}{\ot{1}}=\oz{2}+\cO(z_o^2)$. Inverting these we obtain
\begin{eqnarray}
\oz{1}(\ott{1})&=&1-e^{-\ott{1}}, \nonumber \\
\oz{2}(\ott{1},\ott{2})&=&\ott{2}+\quart\ott{1}\ott{2}+\frac{1}{192}\ott{1}^2\ott{2}-\frac{1}{256}\ott{1}^3\ott{2}+\cO({\tilde{t_o}}^5),
\end{eqnarray}
which together form the mirror map at the orbifold point in moduli space.\\
Transforming the Yukawa couplings, the Christoffel symbols and the holomorphic ambiguities $\tilde{f}$ to the orbifold coordinates we obtain the propagator around this locus. In the choice of our coordinates (\ref{orbicoord}) the propagator takes the following simple form $S^{11}=S^{12}=S^{21}=0$ and
$$ S^{22}=\frac{1}{16}(\ot{2}^2-\ot{1}^2)+\frac{1}{6144}(\ot{1}^4-6\ot{1}^2\ot{2}^2+5\ot{2}^4)+\cO(t_o^5). $$
In order to match the matrix model expansion one has to choose appropriate coordinates. As explained in \cite{Aganagic:2002wv} the right variables $S_1,S_2$ that match the 't Hooft parameters on the matrix model side are given by
\begin{equation}
S_1=\quart(\ot{1}+\ot{2}),\;S_2=\quart(\ot{1}-\ot{2}).
\end{equation}
In addition the overall normalization of the all genus partition function $\F=\sum_g g_s^{2g-2}\F_g$ has to be determined. By comparing to the matrix model one gets, that the string coupling on the topological side, $g_s^{\rm top}$, is related to the coupling on the matrix model side, $\hat g_s$, by the identification $g_s^{\rm top}=2i\hat g_s$. Using these expressions we find
\begin{equation}
\begin{split}
\F^{orb}_2&=-\frac{1}{240}\left(\frac{1}{S_1^2}+\frac{1}{S_2^2}\right)+\frac{1}{360}-\frac{1}{57600}(S_1^2+60S_1S_2+S_2^2)+\cO(S^4)\\
\F^{orb}_3&=\frac{1}{1008}\left(\frac{1}{S_1^4}+\frac{1}{S_2^4}\right)+\frac{1}{22680}+\frac{1}{34836480}(S_1^2-252S_1S_2+S_2^2)+\cO(S^4)\\
\F^{orb}_4&=-\frac{1}{1440}\left(\frac{1}{S_1^6}+\frac{1}{S_2^6}\right)+\frac{1}{340200}-\frac{1}{82944000}(S_1^2+102S_1S_2+S_2^2)+\cO(S^4).
\end{split}
\end{equation}
The genus two results are in accord with \cite{Aganagic:2002wv}, genus three corrects the misprints in this article and genus four is a prediction on the matrix model.

\subsection{Relation to the family of elliptic curves}
At the beginning of this section we pointed out, that $H(x,y)=0$ defines a family of elliptic curves $\Sigma(z_1,z_2)$ whose $j$-function is given by
\begin{equation}
j(z_1,z_2)=\frac{((1-4z_1-4z_2)^2-48z_1z_2)^3}{z_1^2z_2^2(1-8(z_1+z_2)+16(z_1-z_2)^2)}.
\end{equation}
Using the usual $j$-function description (\ref{jfunct}) one can establish a relation between the elliptic parameter $q=e^{2\pi i\tau}$ and the complex structure variables $z_1$ and $z_2$ which reads
\begin{equation}
q=z_1^2z_2^2+16z_1^3z_2^2+160z_1^4z_2^2+16z_1^2z_2^3+400z_1^3z_2^3+160z_1^2z_2^4+\cO(z^7).
\end{equation}
We observe that
\begin{equation}
\tau=4\d_{\xT{2}}\d_{\xT{2}}\F_0,\quad \d_{\xT{2}}\tau=-4C_{\xT{2}\xT{2}\xT{2}},
\end{equation}
where $\xT{i}$ is obtained from $Q_i^x=e^{2\pi i\xT{i}}$, which hints at that the not instanton corrected parameter $x_1$ or $Q^x_1$, respectively, is merely an auxiliary parameter.\\
\cite{ABK} work with an isogenous description of $\S(z_1,z_2)$. They use the Segre embedding of $\P^1\times\P^1$ into $\P^3$ given by the map
\begin{equation}
([x_0:x_1],[x'_0:x'_1])\mapsto[X_0:X_1:X_2:X_3]=[x_0x'_0,x_1x'_0,x_0x'_1,x_1x'_1],
\end{equation}
where $[x_0:x_1]$ and $[x'_0:x'_1]$ are homogeneous coordinates of the $\P^1$'s and $X_0,\dots,X_3$ are homogeneous coordinates of $\P^3$. Then $\tilde{\S}(\tilde{z}_1,\tilde{z}_2)$ is given by the complete intersection of $\P^1\times\P^1$, defined by $X_0X_3-X_1X_2$, with the hypersurface given by $X_0^2+\tilde{z}_1X_1^2+X_2^2+\tilde{z}_2X_3^2+X_0X_3$. Its $j$-function reads
\begin{equation}
\tilde{\jmath}(\tilde{z}_1,\tilde{z}_2)=\frac{((1-4\tilde{z}_1-4\tilde{z}_2)^2+192\tilde{z}_1\tilde{z}_2)^3}{\tilde{z}_1\tilde{z}_2(1-8(\tilde{z}_1+\tilde{z}_2)+16(\tilde{z}_1-\tilde{z}_2)^2)^2}.
\end{equation}
Defining $\tilde{q}=e^{2\pi i\tilde{\tau}}$ we can calculate that $\tilde{\tau}=\d_{\xT{2}}\d_{\xT{2}}\F_0$, i.e. their modular parameters are related by a simple rescaling by a factor of 4
\begin{equation}
\tau=4\tilde{\tau}.
\end{equation}
This transfers to a rescaling of the periods of the elliptic curve, similar to the discussion in section \ref{modexpP2}.

With this input it is possible to write the full non-holomorphic $F_1$ as
\begin{equation}
F_1=-\log \sqrt{\tilde{\tau}_2}\eta(\tilde{\tau})\bar{\eta}(\bar{\tilde{\tau}}).
\end{equation}

\section{$\K_{\HirzeF_1}=\cO(-2,-3) \rightarrow \HirzeF_1$}
\label{localF1} 
We are considering the non-compact Calabi-Yau geometry $\cO(-2,-3)\rightarrow\HirzeF_1$, i.e.~the canonical line bundle over the Hirzebruch surface $\HirzeF_1=\B\P^2_1$, where $\B\P^2_1$ denotes the first del Pezzo surface, i.e. $\P^2$ with one blow up. This local model can be obtained again from the compact elliptic fibration over $\HirzeF_1$ with fiber $X_6(1,2,3)$. The three complexified K\"ahler volumes have the corresponding Mori cone generators $(-6;3,2,1,0,0,0,0),(0;0,0,-1,1,-1,1,0),(0;0,0,-2,0,1,0,1)$.\\
A Picard-Fuchs system governing the periods of the global mirror geometry is given by
\begin{equation}
\begin{split}
&\cD_1=\Theta_1(\Theta_1-2\Theta_2-\Theta_3)-18z_1(1+6\Theta_1)(5+6\Theta_1)\\
&\cD_2=\Theta_2(\Theta_2-\Theta_3)-z_2(-1+\Theta_1-2\Theta_2-\Theta_3)(\Theta_1-2\Theta_2-\Theta_3)\\
&\cD_3=\Theta_3^2-z_3(\Theta_1-2\Theta_2-\Theta_3)(\Theta_2-\Theta_3).
\end{split}
\end{equation}

Now let us turn to the non-compact geometry. The toric data of local $\HirzeF_1$ is summarized in the following matrix
\begin{equation}
(V|Q)=\left(\begin{array}{ccc|cc} 0&0&1&-2&-1\\1&0&1&1&0\\ -1&-1&1&0&1\\ -1&0&1&1&-1 \\ 0&1&1&0&1 \end{array}\right).
\end{equation}
From there we conclude the following quantities\footnote{~ Using toric geometry it is only possible to determine an one-parameter family of classical intersection numbers $C^{(0)}_{ijk}$, resulting in an one-parameter family for $\int_M c_2J_i$. Their correct values are fixed by a limiting procedure of local $\HirzeF_1=\B\P^2_1$ to local $\P^2$ which is described below.}
\begin{equation} \label{topdataF1}
\begin{array}{ll}
a) & Q^1=(-2,1,0,1,0),\; Q^2=(-1,0,1,-1,1) \\
b) & Z=\{x_1=x_3=0\}\cup\{x_2=x_4=0\} \\
c) & M=(\C^5[x_0,\cdots,x_4]\setminus Z)/(\C^*)^2 \\
d) & H(x,y)=y^2-x^3-(1-4z_1)x^2+8z_1z_2x-16z_1^2z_2^2 \\
e) & \cD_1=\Theta_1(\Theta_1-\Theta_2)-z_1(2\Theta_1+\Theta_2)(1+2\Theta_1+2\Theta_2)\\
   & \cD_2=\Theta_2^2-z_2(\Theta_2-\Theta_1)(2\Theta_1+\Theta_2)\\
   & \Delta=(1-4z_1)^2-z_2(1-36z_1+27z_1z_2)\\
f) & C^{(0)}_{111}=-\frac{1}{3},\; C^{(0)}_{112}=-\frac{1}{3},\; C^{(0)}_{122}=-\frac{1}{3},\; C^{(0)}_{222}=\frac{2}{3}\\
g) & \int\limits_M c_2J_1=-2,\; \int\limits_M c_2J_2=0.
\end{array}
\end{equation}
$H(x,y)=0$ defines a family of elliptic curves $\S(z_1,z_2)$ whose $j$-function is given by
\begin{equation}\label{jfctF1}
j(z_1,z_2)=\frac{((1-4z_1)^2+24z_1z_2)^3}{z_1^3z_2^2((1-4z_1)^2-z_2(1-36z_1+27z_1z_2))}.
\end{equation}

\subsection{Solving the topological string on local $\HirzeF_1$ at large radius}
The mirror map at the point of large radius is given by
\begin{equation}
\begin{array}{l}
  2 \pi i T_1(z_1,z_2) = \log{z_1} + 2 z_1 + 3 z_1^2 - 4 z_1 z_2 + \frac{20}{3} z_1^3 + 24 z_1^2 z_2
                         + O(z^4) \\
  2 \pi i T_2(z_1,z_2) = \log{z_2} + z_1 + \frac{3}{2} z_1^2 -2 z_1 z_2 + \frac{10}{3} z_1^3 + 
                         - 12 z_1^2 z_2 + \cO(z^4).\\
\end{array}
\end{equation}
Inverting the series we obtain for $Q_i = e^{2 \pi i T_i}$
\begin{equation}
\begin{array}{l}
  z_1(Q_1,Q_2) = Q_1 - 2 Q_1^2 + 3 Q_1^3 + 4 Q_1^2 Q_2 -4(Q_1^4 + Q_1^3 Q_2) + \cO(Q^5)\\
  z_2(Q_1,Q_2) = Q_2 - Q_1 Q_2 + Q_1^2 Q_2 +2 Q_1 Q_2^2 - Q_1^3 Q_2 + \cO(Q^5).\\
\end{array}
\end{equation}
Now, one realizes again that there is a relation between the $Q$ coordinates:
\begin{equation}
  \frac{Q_1}{Q_2^2} = \frac{z_1}{z_2^2} = e^{2 \pi i (T_1 - 2T_2)} =: Q^x_1.
\end{equation}
Defining further $Q^x_2 := Q_2$ and $x_1 = \frac{z_1}{z_2^2}$ as well as $x_2 = z_2$ one finds that 
\begin{equation}
\begin{array}{l}
  x_1(Q^x_1,Q^x_2) = Q^x_1, \\
  x_2(Q^x_1,Q^x_2) = Q^x_2 - Q^x_1 {Q^x_2}^3 + 2 Q^x_1 {Q^x_2}^4 + \cO(Q^6).\\
\end{array}
\end{equation}
The Yukawa couplings can be fixed through the relation $\d_{T_i} \d_{T_j} \d_{T_k} \F_0 = C_{T_i T_j T_k}$ and the known genus zero GV invariants up to a dependence on one unfixed parameter. This unfixed parameter can be determined by the fact that there exists a limit of local $\HirzeF_1$ to local $\P^2$, as $\HirzeF_1=\B\P^2_1$. This blow-down limit can be seen by comparing the two $j$-functions (\ref{jfctF1}),(\ref{j1}) and turns out to be
\begin{equation*}
z_1\rightarrow0,\text{ with } z_1z_2=z\text{ fixed.}
\end{equation*}
We obtain the following Yukawa couplings
\begin{eqnarray}
C_{111}=\frac{-1-4z_1^2+z_2-z_1(7-6z_2)}{3z_1^3\Delta},\quad
C_{112}=\frac{-1+8z_1^2+z_2+z_1(2-3z_2)}{3z_1^2z_2\Delta},\nonumber\\
C_{122}=\frac{z_2(1-12z_1)-(1-4z_1)^2}{3z_1z_2^2\Delta},\quad
C_{222}=\frac{2(1-4z_1)^2+z_2(1-60z_1)}{3z_2^3\Delta}.
\end{eqnarray}

The next step is to determine the propagators of local $\HirzeF_1$. This is best done in $x$ coordinates, where one finds again that some Christoffel symbols are either trivial or have a rational form
\begin{equation}
\Gamma^1_{11} = -\frac{1}{x_1}, \quad \Gamma^1_{12}=0, \quad \Gamma^1_{22} = 0.
\end{equation}
Choosing $\tilde{f}^1_{11} = -\frac{1}{x_1},~ \tilde{f}^1_{12}=0,~ \tilde{f}^1_{21}=0,~ \tilde{f}^1_{22}=0$, one
finds from (\ref{PropEq}) that $S^{11}$, $S^{12}$ are immediately zero.
Demanding symmetry we are able to fix all ambiguities $\tilde{f}^i_{jk}$ by the choice
\begin{equation}
\begin{split}
\tilde{f}^1_{11}&=-\frac{1}{x_1},\quad
\tilde{f}^2_{11}=-\frac{x_2}{12x_1^2\Delta_x}(1-x_2-12x_1x_2^2+49x_1x_2^3-36x_1x_2^4+32x_1^2x_2^4-12x_1^2x_2^5),\\
\tilde{f}^2_{12}&=-\frac{1}{12x_1\Delta_x}(3-3x_2-32x_1x_2^2+144x_1x_2^3-108x_1x_2^4+80x_1^2x_2^4),\\
\tilde{f}^2_{22}&=-\frac{1}{12x_2\Delta_x}(20-21x_2-176x_1x_2^2+828x_1x_2^3-648x_1x_2^4+384x_1^2x_2^4),
\end{split}
\end{equation}
where $\Delta_x$ denotes the discriminant in $x$ coordinates and all other combinations of $\tilde{f}^i_{jk}$ are either zero or follow by symmetry. This singles out one non-vanishing propagator only, given by $S^{22}(x_1,x_2)=\frac{x_2^2}{12}-\frac{1}{3}x_1x_2^4+x_1x_2^5+4x_1^2x_2^7+\cO(x^{10})$. After tensor transforming to $z$ coordinates we obtain
\begin{equation}
S^{ij}=\begin{pmatrix}
S(z_1,z_2) & \displaystyle\frac{z_2}{2 z_1}\,S(z_1,z_2) \\
\displaystyle\frac{z_2}{2 z_1}\,S(z_1,z_2) & \displaystyle\frac{z_2^2}{4 z_1^2}\,S(z_1,z_2)
       \end{pmatrix},
\end{equation}
where $S(z_1,z_2)=\frac{z_1^2}{3}-\frac{4z_1^3}{3}+4z_1^3z_2+16z_1^4z_2+\cO(z^6)$. This again has a similar form as in the case of local $\HirzeF_0$.\\
In addition, we fix the holomorphic ambiguity of the covariant derivative of $S^{ij}$, (\ref{DS}), and obtain, that in $x$ coordinates there are two non-zero contributions only, given by
\begin{equation}
\begin{split}
f_1^{22}&=-\frac{x_2^2}{144x_1\Delta_x}(3-3x_2+4x_1x_2^2)(1-8x_1x_2^2+24x_1x_2^3+16x_1^2x_2^4),\\
f_2^{22}&=-\frac{x_2}{144\Delta_x}(8-9x_2)(1-8x_1x_2^2+24x_1x_2^3+16x_1^2x_2^4).
\end{split}
\end{equation}
The $f_i^{jk}$ in $z$ coordinates are again obtained after tensor transformation. 

Further we can express the covariant derivative of $F_1$ through the generator $S$ by
\begin{equation}
D_iF_1=\frac{1}{2}C_{ijk}S^{jk}+A_i.
\end{equation}
As the free energy function of genus one is given by
\begin{equation}
\begin{split}
F_1&=\log\left(\Delta^{-\frac{1}{12}}z_1^{-\frac{7}{12}}z_2^{-\frac{1}{2}}\det (G_{i\bar\jmath}))^{-\frac{1}{2}}\right),\\
\F_1(T_1,T_2)&=-\frac{1}{12}\log(Q_1) - \frac{1}{12}(2Q_1 + Q_2) - \frac{1}{24}(2Q_1^2  +  6Q_1 Q_2 + Q_2^2) + \cO(Q^3),
\end{split}
\end{equation}
we find that $A_i=\d_i A$ and
\begin{equation}\label{F1A}
A=-\frac{1}{24}\log\Delta + \frac{1}{24}\log z_1 + \frac{1}{12}\log z_2.
\end{equation}
Now, we are prepared to perform the direct integration procedure. Demanding the gap at the conifold and using further the known constant map contributions we are able to fix the ambiguities up to genus three. In this more general two parameter model with one discriminant component of degree three the number of coefficients in $A_g$ is
\begin{equation}
\binom{(2g-2)\deg\Delta+2}{2}=10 - 27 g + 18 g^2,
\end{equation}
i.e. at genus three we have to fix 91 coefficients in the holomorphic ambiguity.\\
The invariants can be found in the appendix \ref{GVInvariants}. The solutions around the conifold locus are described in the next section.

\subsection{Solving the topological string on local $\HirzeF_1$ at the conifold locus}
In order to apply the gap condition in this example, we have to transform and solve the Picard-Fuchs system at a specific point on the conifold locus. We make the choice $z_1=2$, $z_2=-\frac{1}{2}$. Again we define two variables which vanish at this point
\begin{equation}
  z_{c,1} = 1- \frac{z_2}{-\frac{1}{4}(z_1-2)-\frac{1}{2}}, ~~~z_{c,2}=1-\frac{z_2}{4 (z_1-2)-\frac{1}{2}}.
\end{equation}
$z_{c,1}$ is a coordinate normal to the conifold divisor and $z_{c,2}$ describes a tangential direction. Transforming the Picard-Fuchs system to these
coordinates we find the following set of periods:
\begin{eqnarray}
  \omega^c_0 & = & 1, \nonumber \\
  \omega^c_1 & = & z_{c,1} + \frac{6773 z_{c,1}^2}{14450} - \frac{58 z_{c,1} z_{c,2}}{7225} 
                 - \frac{z_{c,2}^2}{1445} + \cO(z_c^3), \nonumber \\
  \omega^c_2 & = & z_{c,2} + \frac{10858 z_{c,1}^2}{7225} + \frac{2871 z_{c,2}^2}{2890} 
               - \frac{4886 z_{c,1} z_{c,2}}{7225} + \cO(z_c^3).
\end{eqnarray}
Next, we can express the $z_{c,i}$ through the mirror coordinates $t_{c,1}:=\omega^c_1$ and $t_{c,2} := \omega^c_2$ by inverting the above series
\begin{eqnarray}
  z_{c,1}(t_{c,1},t_{c,2}) & = & t_{c,1} - \frac{6773 t_{c,1}^2}{14450} + \frac{58 t_{c,1} t_{c,2}}{7225}
                                 + \frac{t_{c,2}^2}{1445} + \cO(t_c^3), \nonumber \\
  z_{c,2}(t_{c,1},t_{c,2}) & = & t_{c,2} - \frac{10858 t_{c,1}^2}{7225} + \frac{4886 t_{c,1} t_{c,2}}{7225} 
                                 - \frac{2871 t_{c,2}^2}{2890} + \cO(t_c^3).
\end{eqnarray}
Transforming the Yukawa couplings, the Christoffel symbols and the holomorphic ambiguities $\tilde{f}$ to the conifold coordinates we obtain the propagator around this locus. In the choice of our coordinates the propagator takes the following form
\begin{equation}
\begin{split}
S^{11}&=\frac{5}{12}-\frac{2\ct{1}}{25}-\frac{337\ct{1}^2}{10625}-\frac{4\ct{1}\ct{2}}{2125}+\cO(t_c^3),\\
S^{12}&=-\frac{55}{4}+\frac{66\ct{1}}{25}+\frac{11121\ct{1}^2}{10625}+\frac{132\ct{1}\ct{2}}{2125}+\cO(t_c^3),\\
S^{22}&=\frac{1815}{4}-\frac{2178\ct{1}}{25}-\frac{366993\ct{1}^2}{10625}-\frac{4356\ct{1}\ct{2}}{2125}+\cO(t_c^3).
\end{split}
\end{equation}
Again the gap condition in combination with the known leading behavior at the large radius point suffices to fix all coefficients in the holomorphic ambiguity. From the conifold alone we get at genus two 27 out of 28 unknowns and at genus three 90 out of 91 unknowns. Our results read
\begin{equation}
\begin{split}
\F_2^c&=\frac{1}{48\ct{1}^2}+\frac{1567}{9000000}+\frac{98333}{1593750000}\ct{1}-\frac{123}{10625000}\ct{2}+\cO(t_c^2)\\
\F_3^c& =\frac{25}{1008\ct{1}^4}+\frac{480217}{283500000000}+\frac{106245283\ct{1}}{17929687500000}+\frac{69949\ct{2}}{167343750000}+\cO(t_c^2).
\end{split}
\end{equation}

\subsection{Relation to the family of elliptic curves}
Starting point is again the $j$-function of $\S(z_1,z_2)$ which we will repeat here
\begin{equation}
j(z_1,z_2)=\frac{((1-4z_1)^2+24z_1z_2)^3}{z_1^3z_2^2((1-4z_1)^2-z_2(1-36z_1+27z_1z_2))}.
\end{equation}
Using again the usual $j$-function description (\ref{jfunct}) one can establish a relation between the elliptic parameter $q=e^{2\pi i\tau}$ and the complex structure variables $z_1$ and $z_2$ which reads
\begin{equation}
q=z_1^3z_2^2+16z_1^4z_2^2+160z_1^5z_2^2-z_1^3z_2^3-60z_1^4z_2^3+\cO(z^8).
\end{equation}
We observe that
\begin{equation}
\tau=\d_{\xT{2}}\d_{\xT{2}}F_0,\quad \d_{\xT{2}}\tau=-C_{\xT{2}\xT{2}\xT{2}},
\end{equation}
where $\xT{i}$ is obtained from $Q_i^x=e^{2\pi i\xT{i}}$, which hints at that the not instanton corrected parameter $x_1$ or $Q^x_1$, respectively, is merely an auxiliary parameter.
As in the previous cases it is possible to write the full non-holomorphic $F_1$ as
\begin{equation}
F_1=-\log \sqrt{\tau_2}\eta(\tau)\bar{\eta}(\bar{\tau})+A,
\end{equation}
where $A$ is given by (\ref{F1A}).

\section{Summary and further directions}
\label{conclusion} 

In this article we find convincing  evidence that closed 
topological string theories on non-compact Calabi-Yau spaces 
whose mirror can be reduced to Riemann surfaces is completely integrable 
using the holomorphic anomaly equation and the gap at the divisors 
at which a single cycle vanishes. The physical argument for the gap 
from the local form of the effective action in the presence of 
a single black hole hypermultiplet state that becomes 
massless at the nodal singularity~\cite{HKQ} applies also 
after the decompactification limit. The massless hypermultiplet 
is now a dyonic hypermultiplet of a rigid 4d theory. This extends 
in particular to the geometric engineering limits, which leads 
to $N=2$ supersymmetric gauge theories in $4d$. Indeed the gap was found 
in simple Seiberg-Witten theories~\cite{HK} and it  made 
the holomorphic anomaly equations integrable in these cases. 

Generally there are two sorts of parameters associated to the 
geometry $(\Sigma_g,\lambda)$. There are $r$ parameters, which are 
given by periods over $H^1(\Sigma_g)$. The monodromy acts on them 
and $T$ duality requires that their occurrence in higher genus 
amplitudes is organized in terms of almost holomorphic modular 
forms, which correspond to non-trivial components of the 
propagators $S^{ij}$. Further there might be $m$ parameters encoding 
the non-vanishing residua of the meromorphic form $\lambda$. 
The monodromy acts trivially on them. In mathematics they are referred to as 
isomonodromic deformations. We find that they occur in rational 
expressions in the amplitudes.   

In Seiberg-Witten theory the $r$ parameters correspond to the number 
of $U(1)$ vector multiplets in the Coulomb phase, while $m$ 
parameters are the masses of perturbative hypermultiplets.     
Similar del Pezzo surfaces with $1+m$ K\"ahler parameters have 
genus one mirror curves and we could identify the one parameter 
that corresponds to an integral over $H^1(\Sigma_1)$ and the 
$m$ residue parameters by choosing a parameterization in 
which we have only one non-trivial propagator. In all 
cases we found by a local analysis of the gap condition 
near the discriminant components with single vanishing cycle 
that there are sufficiently many conditions to solve the theory.    
For Seiberg-Witten  theories with matter fields  
this has been established in~\cite{HKprogress}.

In recent years strong relations between  
topological string theory on local Calabi-Yau manifolds
and matrix models and other integrable structures such as 
Chern-Simons theory have been discovered. These developments 
have been excellently reviewed in~\cite{Marino:2005sj}\cite{Marino:2004zz}.

In particular~\cite{Chekhov:2005kd}\cite{Dijkgraaf:2002fc} show 
that rigid special geometry, which is essential 
in making the ring of the propagators close under 
derivatives (section \ref{directintegration}), is an 
intrinsic property of the multi cut matrix model if 
the filling fractions are considered as parameters. 
Further it was argued in~\cite{Eynard:2007hf} that 
the  method of solving the recursive loop equation 
using the Bergman kernel and the kernel differentials 
of~\cite{Eynard:2007kz} can be made modular by adding 
a non-holomorphic modular completion to the Bergmann 
kernel. It was further shown in~\cite{Eynard:2007hf} that 
this completion makes the formalism of~\cite{Eynard:2007kz} 
compatible with the holomorphic anomaly equation. The modular 
property has not yet been derived within the matrix model. 
In fact the analysis of~\cite{Eynard:2007hf} is inspired by the 
way modularity is realized in the higher genus expansion of 
topological string  theory on non-compact Calabi-Yau and 
Seiberg-Witten theory~\cite{ABK}\cite{HK}, where $T$ 
or $S$ duality is an intrinsic property. In any case it is 
clear that the matrix model correlation functions in the 
$\frac{1}{N^2}$ expansions fulfill the holomorphic anomaly 
equations. Moreover~\cite{Marino:2006hs} applies the formalism 
of~\cite{Eynard:2007kz} to local mirror curves and 
successfully checks expansions of closed and open 
low genus amplitudes large against A-model calculations.
This leads to the expectation that the $F_g$ for many multi-cut 
matrix  models are solvable using the modular properties of 
the spectral curve and the gap condition.

To summarize we have good evidence that the holomorphic anomaly equation 
and the gap conditions solve the closed amplitudes for 
the following cases: non-compact Calabi-Yau with mirror curves, 
Seiberg-Witten theories and for many multi cut matrix 
models. What makes the claim plausible in general 
is that the Riemann surfaces have in the co-dimension 
one locus in the moduli space just one type of degeneration, 
the nodal degeneration, which exhibits as local property 
the gap behavior. E.g.~$SU(N)$ theories can be degenerated to 
$SU(N_1)\times\ldots\times SU(N_k)$ theories, with $\sum_{i=1}^k N_i=N$ 
by stretching  higher genus components of the curve apart. 
Such operations can not affect the local leading behavior 
of $F_g$ near the pinching cycles and for $N_i=2$ the gap 
is established \cite{HK}. 

Due to a more extensive use of the symmetry the method 
outlined here is more efficient then any other to calculate 
the $F_g$ for high $g$ and provides global expressions instead of local 
expansions. Combined with numerical analysis of asymptotic 
expansions this has applications in investigations 
of non-perturbative completions of topological string 
theory~\cite{Marino:2008ya}\cite{Eynard:2008yb}. 
Understanding the role of holomorphicity and modularity, 
which are the basis of our approach, could give decisive 
hints for such completions.

One might further speculate that the approach extends to open strings. 
The open string version of the holomorphic anomaly equation 
in the presence of open string moduli has yet some problems~\footnote{~We thank 
Marcos Mari\~no for a discussion on the issue.}~\cite{Eynard:2007hf}.  
The open string variables are not subject to modular transformations 
and in this sense similar to the $m$ residue parameters. But in the open case we have 
so far not understood how to provide enough boundary conditions 
to make the holomorphic anomaly approach completely integrable. 
For the open string on compact Calabi-Yau spaces  without open string moduli 
no particular structure  has been found at the boundary of the closed string
moduli space~\cite{Walcher:2007tp}.

Extracting the full constraints from the local 
analysis of the multi parameter gap condition is also relevant 
to multi parameter global Calabi-Yau spaces and could 
lead to integrability of these systems. Different then in the 
one parameter cases where the situation has been analyzed in
\cite{HKQ}\cite{Hosono:2007vf}\cite{Haghighat:2008ut} one can employ here 
further known limits such as the large base limit in K3 fibrations, in 
which formulas for the all genus generating functions of GW invariants 
have been  mathematically rigorously 
established in~\cite{KMPS}.

\subsection*{Acknowledgment}
We like to thank Vincent Bouchard, Minxin Huang, Thomas Grimm, 
Rainald Flume, Marcos Mari\~no, Sara Pasquetti, Piotr Sulkowski 
and S.T.~Yau for discussions.\\
The work of  B.H.~and M.R.~is supported by the German Excellence 
Initiative via the graduate school BCGS. A.K.~is supported by 
DOE grant DE-FG02-95ER40896.

\newpage
\appendix
\section{Modular anomaly versus holomorphic anomaly}
\label{modularity} 
Physically the amplitudes $F_g$ of the topological string are invariant under
the space-time modular group $\Gamma$ of the target space. This is the most important 
restriction on these functions. The nicest case is when the B-model geometry is a 
family of elliptic curves. Then $\Gamma$ is a subgroup of ${\rm SL}(2,\Z)$ 
and the classical theory of modular forms applies. We will recapitulate below  
the relevant aspects of ${\rm SL}(2,\Z)$ almost holomorphic modular forms. 
This gives some insight in the interplay between the breaking of the
modularity and the breaking of holomorphicity. The different modular forms
that we need for the general families of elliptic 
curves, i.e. general two cut matrix models, follow from the Picard-Fuchs equations. 
The relation between the Picard-Fuchs equations and modular forms is again a 
classical subject, which has been beautifully reviewed in \cite{zagierreview}.

\subsection{${\rm PSL}(2,\Z)$ modular forms} 

We define $q:=e^{2 \pi i \tau}$, with $\tau\in \H_+=\{\tau \in
\C\,|\,{\rm Im}(\tau)=\frac{1}{2 i}(\tau-\bar \tau)> 0\}$ 
and the projective action  ${\rm PSL}(2,\Z)$ of
$\Gamma_1={\rm SL}(2,\Z)=\left\{\gamma=\left(\begin{array}{cc} a \ & b \\ c &
      d\end{array}\right)\,|\,ad - bc =1,\  a,b,c,d \in \Z\right\}$
on $\H_+$  by
\begin{equation}
\tau \mapsto \tau_\gamma=\frac{a\tau + b}{c \tau + d},
\end{equation}
for $\gamma\in \Gamma_1$. It follows that 
\begin{equation} 
\frac{1}{{\rm Im}(\tau_\gamma)}=\frac{ (c\tau +d)^2}{\itau}-2 i c (c \tau
+d)=\frac{|c \tau +d |^2}{\itau}\ .
\label{imtautrans}
\end{equation}  
Modular forms  of $\Gamma_1$ transform  as 
\begin{equation}
f_k(\tau_\gamma)=(c \tau + d)^k f_k(\tau)
\label{formtransform}
\end{equation}
with weight $k\in \Z$ for all $\tau\in \H_+$ and $\gamma\in
\Gamma_1$, are meromorphic for $\tau\in \H_+$  and grow 
like ${\cal O}(e^{C {\rm Im}(\tau)})$ for ${\rm Im}(\tau)\rightarrow \infty$ and 
${\cal O}(e^{C/{\rm Im}(\tau)})$ for ${\rm Im}(\tau)\rightarrow 0$  with $C>0$.  
 A strategy to build modular forms of 
weight $k$ is to sum over orbits of $\Gamma_1$  
\begin{equation} 
G_k=\frac{1}{2} \sum_{m,n\in \Z\atop (m,n)\neq (0,0)} \frac{1}{(m \tau
  +n)^k}\ .
\label{Eisenstein}  
\end{equation}
It is easy to see that this expression transforms like (\ref{formtransform}), 
converges absolutely for $k>2$ and vanishes for $k$ odd. In the standard
definition of the Eisenstein series $E_k$ the sum runs over coprime $(m,n)$, 
which yields a proportionality $G_k(\tau)=\zeta(k) E_k(\tau)$, where
$\zeta(k)=\sum_{n\ge 1}\frac{1}{n^k}$. One shows (\cite{zagierreview}) the
central fact that $E_4,E_6$ (or $G_4$,$G_6$ of course)  generate freely 
the graded (by $k$) ring of modular forms ${\cal M}_*(\Gamma_1)$. 

Still one may spot two shortcomings. Firstly the ring ${\cal M}_*(\Gamma_1)$ 
does not close under any  differentiation and secondly there should be a 
modular form for weight $2$. 
These facts are related as $d_\tau=\frac{d}{2 \pi i d\tau}$ has weight $2$.
The second is remedied by an $\epsilon$ regularization in the sum 
$G_{2,\epsilon}=\frac{1}{2} \sum_{m,n\in \Z\atop (m,n)\neq (0,0)} \frac{1}{(m
  \tau +n)^k |m \tau +n|^\epsilon}$ after which it is possible to 
define $G_2=\lim _{\epsilon\rightarrow 0} G_{2,\epsilon}$. Then all $G_k$,
$k\in 2 \Z$, $k\ge 2$ have a Fourier expansion\footnote{~Note that the
Eisenstein series start with coefficient $1$.} in $q=\exp(2 \pi i \tau)$
\begin{equation} 
G_k(\tau)=\frac{(2 \pi i)^k}{(k-1)!} \left(- \frac{B_k}{2 k}+\sum_{n=1}^\infty
  \sigma_{k-1}(n) q^n\right),
\label{ResumEisenstein}
\end{equation}
with $\sigma_k(n)=\sum_{p|n} p^k$ the sum of $k$th powers of positive divisors
of $n$ and $\sum_{k=0}^\infty \frac{B_k x^k}{k!}=\frac{x}{e^x-1}$ defining the
Bernoulli numbers $B_k$, e.g. $B_2=\frac{1}{6}$, $B_4=-\frac{1}{30}$,
$B_6=\frac{1}{42}$, $B_8=-\frac{1}{30}$, $B_{10}=\frac{5}{66}$,
$B_{12}=-\frac{691}{2730}$, $B_{14}=\frac{7}{6}$ etc. 

Very much like in QFT the regularization introduces an anomaly in the
symmetry transformation so that $E_2$ transforms
\begin{equation} 
E_2(\tau_\gamma)=(c \tau + d)^2 E_2(\tau)-\frac{ 6 i c}{\pi} (c \tau+d)
\label{E2trans}
\end{equation}
with an inhomogeneous term. 

At least $(E_2,E_4,E_6)$ form a ring, the ring of quasi modular 
holomorphic forms ${\cal M}^!$, which closes under differentiation, i.e.
\begin{equation}
d_{\tau} E_2=\frac{1}{12} (E_2^2 - E_4),\ \ d_{\tau} E_4=\frac{1}{3} (E_2 E_4
- E_6),\ \ d_{\tau} E_6=\frac{1}{2} (E_2 E_6 - E_4^2)\ .
\label{Ekderivatives}
\end{equation}  
Using (\ref{imtautrans}) and (\ref{E2trans}) we see that the inhomogeneous
terms in (\ref{imtautrans},\ref{E2trans}) cancel so that
\begin{equation} 
\hat E_2(\tau)=E_2(\tau)-\frac{3}{\pi {\rm Im}(\tau)}
\label{E2hat}
\end{equation}
transforms like a modular form of weight $2$, albeit not a holomorphic one. 
$(\hat E_2,E_4,E_6)$ form the ring of almost holomorphic modular  
forms of $\Gamma_1$. The latter closes under the Maass derivative, 
which acts on forms of weight $k$ by 
\begin{equation}
D_{\tau} f_k=\left(d_\tau-\frac{k}{4 \pi {\rm Im}(\tau)}\right)f_k\   
\label{Massderivatives}
\end{equation}  
and maps $D_\tau: {\cal M}^!_k \rightarrow {\cal M}^!_{k+2}$. Note that the
equations (\ref{Ekderivatives}) hold with $d_\tau$ replaced by $D_\tau$ 
and $E_2(\tau)$ replaced by $\hat E_2(\tau)$. This Maass derivative 
corresponds to the covariant derivative that appears in topological 
string theory (\ref{HAEq}).  

From the physical point of view there seems the following  
story behind these well known mathematical facts. The holomorphic 
propagator, which can be made proportional to $E_2$, see (\ref{smodular})  
needs some regularization, which breaks $T$ duality. The latter is 
restored by adding the non-holomorphic term (\ref{E2hat}). The 
modular anomaly and the holomorphic anomaly are in this sense 
counterparts, which cannot both be realized at least 
perturbatively. $T$-duality is physically better
motivated. Attempts in the literature, e.g.~in an interesting
paper~\cite{Eynard:2008yb}, to define a holomorphic and 
modular non-perturbative completion by summing over orbits seem to 
make sense only if absolute convergence in the moduli is established, 
which is hard. 

$F_1$ is an index, which is finite for smooth compact spaces. 
It diverges therefore only from singular configurations, that 
occur if e.g.~the discriminant of the elliptic curve given below for 
the Weierstrass form $y^2=4x^3-3 x E_4+E_6$ 
\begin{equation}
\Delta(\tau)=\eta^{24}(\tau)=q \prod_{n=1}^\infty (1-
q^n)^{24}=\frac{1}{1728}(E_4^3(\tau)-E_6^2(\tau))\ , 
\label{Delta}
\end{equation}
vanishes. Note that the $j$ for this curve is 
\begin{equation}
j=1728 \frac{E_4^2}{E_4^3-E_6^2}=\frac{1}{q}+744+196884 q + 21493760 q^2+{\cal
  O}(q^3)\ .
\label{jfunct}
\end{equation} 

It follows from (\ref{formtransform}) that  
$\eta(\tau_\gamma)=(c\tau +d)^\frac{1}{2}\eta(\tau)$ transforms with weight
$\frac{1}{2}$ and from (\ref{Ekderivatives}) that 
\begin{equation} 
d_\tau \log(\eta(\tau))=\frac{1}{24} E_2(\tau) .
\label{derlogeta}
\end{equation} 
Further  from (\ref{imtautrans}) we see that 
$\sqrt{{\rm Im}(\tau)}|\eta(\tau)|^2$ is an almost holomorphic
modular invariant and from (\ref{Ekderivatives},\ref{E2hat},\ref{Delta}) that
\begin{equation} 
d_\tau \log(\sqrt{{\rm Im}(\tau)}|\eta(\tau)|^2 )=\frac{1}{24} \hat E_2(\tau). 
\label{derF1}
\end{equation} 

We need also the theta functions of  general characteristic 
\begin{equation} 
{\theta\left[a \atop b\right](z,\tau)=\sum_{n\in \Z}\exp\left(\pi i
(n+a)\tau (n+a)+ 2 \pi i \sum_i (z+b)n\right)\ .}
\label{gentheta}
\end{equation}

\section{Gopakumar-Vafa invariants of local Calabi-Yau manifolds}
\label{GVInvariants}

\begin{sidewaystable}[h!]
\centering{\tiny{
\begin{tabular}[h]{|c|ccccccccccccc|} 
\hline
$g\backslash d$& 1& 2& 3&4& 5& 6& 7& 8& 9& 10& 11&12& 13\\
\hline
0\!\!\!\!\!\!&\!\!\!\!\!\! 3\!\!\!\!\!\!&\!\!\!\!\!\! -6\!\!\!\!\!\!&\!\!\!\!\!\! 27\!\!\!\!\!\!&\!\!\!\!\!\! -192\!\!\!\!\!\!&\!\!\!\!\!\! 1695\!\!\!\!\!\!&\!\!\!\!\!\! -17064\!\!\!\!\!\!&\!\!\!\!\!\! 188454\!\!\!\!\!\!&\!\!\!\!\!\! -2228160\!\!\!\!\!\!&\!\!\!\!\!\! 27748899\!\!\!\!\!\!&\!\!\!\!\!\! -360012150\!\!\!\!\!\!&\!\!\!\!\!\! 4827935937\!\!\!\!\!\!&\!\!\!\!\!\! -66537713520\!\!\!\!\!\!&\!\!\!\!\!\! 938273463465\\[-.5pt] 
 1\!\!\!\!\!\!&\!\!\!\!\!\! 0\!\!\!\!\!\!&\!\!\!\!\!\! 0\!\!\!\!\!\!&\!\!\!\!\!\! -10\!\!\!\!\!\!&\!\!\!\!\!\! 231\!\!\!\!\!\!&\!\!\!\!\!\! -4452\!\!\!\!\!\!&\!\!\!\!\!\! 80948\!\!\!\!\!\!&\!\!\!\!\!\! -1438086\!\!\!\!\!\!&\!\!\!\!\!\! 25301295\!\!\!\!\!\!&\!\!\!\!\!\! -443384578\!\!\!\!\!\!&\!\!\!\!\!\! 7760515332\!\!\!\!\!\!&\!\!\!\!\!\! -135854179422\!\!\!\!\!\!&\!\!\!\!\!\! 2380305803719\!\!\!\!\!\!&\!\!\!\!\!\! -41756224045650\\[-.5pt] 
 2\!\!\!\!\!\!&\!\!\!\!\!\! 0\!\!\!\!\!\!&\!\!\!\!\!\! 0\!\!\!\!\!\!&\!\!\!\!\!\! 0\!\!\!\!\!\!&\!\!\!\!\!\! -102\!\!\!\!\!\!&\!\!\!\!\!\! 5430\!\!\!\!\!\!&\!\!\!\!\!\! -194022\!\!\!\!\!\!&\!\!\!\!\!\! 5784837\!\!\!\!\!\!&\!\!\!\!\!\! -155322234\!\!\!\!\!\!&\!\!\!\!\!\! 3894455457\!\!\!\!\!\!&\!\!\!\!\!\! -93050366010\!\!\!\!\!\!&\!\!\!\!\!\! 2145146041119\!\!\!\!\!\!&\!\!\!\!\!\! -48109281322212\!\!\!\!\!\!&\!\!\!\!\!\! 1055620386953940\\[-.5pt] 
 3\!\!\!\!\!\!&\!\!\!\!\!\! 0\!\!\!\!\!\!&\!\!\!\!\!\! 0\!\!\!\!\!\!&\!\!\!\!\!\! 0\!\!\!\!\!\!&\!\!\!\!\!\! 15\!\!\!\!\!\!&\!\!\!\!\!\! -3672\!\!\!\!\!\!&\!\!\!\!\!\! 290853\!\!\!\!\!\!&\!\!\!\!\!\! -15363990\!\!\!\!\!\!&\!\!\!\!\!\! 649358826\!\!\!\!\!\!&\!\!\!\!\!\! -23769907110\!\!\!\!\!\!&\!\!\!\!\!\! 786400843911\!\!\!\!\!\!&\!\!\!\!\!\! -24130293606924\!\!\!\!\!\!&\!\!\!\!\!\! 698473748830878\!\!\!\!\!\!&\!\!\!\!\!\! -19298221675559646\\
\hline\end{tabular}}}
\caption{BPS numbers $n^g_d$ of local $\K_{\P^2}$}
\end{sidewaystable} 

\begin{table}[h!]
\centering{\tiny{
\begin{tabular}[h]{|c|ccc|}
\hline
$g\backslash d$\!&\! 14\!&\! 15\!&\!16\\
\hline
 0\!\!\!&\!\!\! -13491638200194\!\!\!&\!\!\! 197287568723655\!\!\!&\!\!\! -2927443754647296\\[-.8pt] 
 1\!\!\!&\!\!\! 733512068799924\!\!\!&\!\!\! -12903696488738656\!\!\!&\!\!\! 227321059950010137\\[-.8pt] 
 2\!\!\!&\!\!\! -22755110195405850\!\!\!&\!\!\! 483361869975894765\!\!\!&\!\!\! -10141562289815822472\\[-.8pt] 
 3\!\!\!&\!\!\! 513289541565539286\!\!\!&\!\!\! -13226687073790872894\!\!\!&\!\!\! 331823525571283260201\\[-.8pt] 
 4\!\!\!&\!\!\! -9179532480730484952\!\!\!&\!\!\! 288379973286696180135\!\!\!&\!\!\! -8687442794831017531164\\[-.8pt] 
 5\!\!\!&\!\!\! 135875843241729533613\!\!\!&\!\!\! -5230662528295888702200\!\!\!&\!\!\! 190039036692844531945431\\[-.8pt] 
 6\!\!\!&\!\!\! -1707886552705077581628\!\!\!&\!\!\! 80979854504456065293006\!\!\!&\!\!\! -3563867463166417898028954\\
\hline
\end{tabular}}}
\caption{BPS numbers $n^g_d$ of local $\K_{\P^2}$}
\end{table}

\begin{table}[!h]
\centering
\begin{tabular}[h]{|c|cccccccc|} 
\hline
      & $d_1$ &   0 &     1 &     2 &     3 &     4 &        5 &       6  \\
\hline
$d_2$ &       &     &       &       &       &        &         &          \\
0     &       &     &    -2 &     0 &     0 &      0 &       0 &       0  \\
1     &       &  -2 &    -4 &    -6 &    -8 &    -10 &     -12 &     -14  \\
2     &       &   0 &    -6 &   -32 &  -110 &   -288 &    -644 &   -1280  \\
3     &       &   0 &    -8 &  -110 &  -756 &  -3556 &  -13072 &  -40338  \\
4     &       &   0 &   -10 &  -288 & -3556 & -27264 & -153324 & -690400  \\
5     &       &   0 &   -12 &  -644 &-13072 &-153324 &-1252040 & -7877210 \\
6     &       &   0 &   -14 & -1280 &-40338 &-690400 &-7877210 &-67008672 \\
\hline
\end{tabular}
\caption{Instanton numbers $n^{g=0}_{d_1d_2}$ of local $\K_{\HirzeF_0}$}
\end{table}

\begin{table}[h!]
\centering
\begin{tabular}[h]{|c|cccccccc|}
\hline
      & $d_1$ & 0     &    1 &   2 &     3 &       4 &        5 &         6   \\
\hline
$d_2$ &       &        &     &     &       &         &          &            \\
0     &       &        &   0 &   0 &     0 &       0 &        0 &         0  \\
1     &       &      0 &   0 &   0 &     0 &       0 &        0 &         0  \\
2     &       &      0 &   0 &   9 &    68 &     300 &      988 &      2698  \\
3     &       &      0 &   0 &  68 &  1016 &    7792 &    41376 &    172124  \\
4     &       &      0 &   0 & 300 &  7792 &   95313 &   760764 &   4552692  \\
5     &       &      0 &   0 & 988 & 41376 &  760764 &  8695048 &  71859628  \\
6     &       &      0 &   0 &2698 &172124 & 4552692 & 71859628 & 795165949  \\
\hline
\end{tabular}
\caption{Genus one GV invariants $n^{g=1}_{d_1d_2}$ of local $\K_{\HirzeF_0}$}
\end{table}

\begin{table}[h!]
\centering
\begin{tabular}[h]{|c|cccccccc|}
\hline
      & $d_1$ & 0     &    1 &   2 &     3 &       4 &        5 &         6   \\
\hline
$d_2$ &       &        &     &     &       &         &          &            \\
0     &       &        &   0 &   0 &     0 &       0 &        0 &         0  \\
1     &       &      0 &   0 &   0 &     0 &       0 &        0 &         0  \\
2     &       &      0 &   0 &   0 &   -12 &    -116 &     -628 &     -2488   \\
3     &       &      0 &   0 & -12 &  -580 &   -8042 &   -64624 &   -371980   \\
4     &       &      0 &   0 &-116 & -8042 & -167936 & -1964440 & -15913228   \\
5     &       &      0 &   0 &-628 &-64624 &-1964440 &-32242268 &-355307838   \\
6     &       &      0 &   0 &-2488&-371980&-15913228&-355307838&-5182075136  \\
\hline
\end{tabular}
\caption{Genus two GV invariants $n^{g=2}_{d_1d_2}$ of local $\K_{\HirzeF_0}$}
\end{table}

\begin{table}[h!]
\centering
\begin{tabular}[h]{|c|cccccccc|}
\hline
      & $d_1$ & 0     &    1 &   2 &     3 &       4 &        5 &         6  \\
\hline
$d_2$ &       &        &     &     &       &         &          &            \\
0     &       &        &   0 &   0 &     0 &       0 &        0 &         0  \\
1     &       &      0 &   0 &   0 &     0 &       0 &        0 &         0  \\
2     &       &      0 &   0 &   0 &     0 &      15 &      176 &      1130  \\
3     &       &      0 &   0 &   0 &   156 &    4680 &    60840 &    501440  \\
4     &       &      0 &   0 &  15 &  4680 &  184056 &  3288688 &  36882969  \\
5     &       &      0 &   0 & 176 & 60840 & 3288688 & 80072160 &1198255524  \\
6     &       &      0 &   0 &1130 &501440 &36882969 &1198255524&23409326968 \\
\hline
\end{tabular}
\caption{Genus three GV invariants $n^{g=3}_{d_1d_2}$ of local $\K_{\HirzeF_0}$}
\end{table}

\begin{table}[h!]
\centering
\begin{tabular}[h]{|c|cccccccc|}
\hline
      & $d_1$ & 0     &    1 &   2 &     3 &       4 &        5 &          6  \\
\hline
$d_2$ &       &        &     &     &       &         &          &             \\
0     &       &        &   0 &   0 &     0 &       0 &        0 &          0  \\
1     &       &      0 &   0 &   0 &     0 &       0 &        0 &          0  \\
2     &       &      0 &   0 &   0 &     0 &       0 &      -18 &       -248 \\
3     &       &      0 &   0 &   0 &   -16 &   -1560 &   -36408 &    -450438 \\
4     &       &      0 &   0 &   0 & -1560 & -133464 & -3839632 &  -61250176 \\
5     &       &      0 &   0 & -18 &-36408 &-3839632 &-144085372&-2989287812 \\
6     &       &      0 &   0 &-248 &-450438&-61250176&-2989287812&-79635105296\\
\hline
\end{tabular}
\caption{Genus four GV invariants $n^{g=4}_{d_1d_2}$ of local $\K_{\HirzeF_0}$}
\end{table}

\begin{table}[h!]
\centering
\begin{tabular}[h]{|c|ccccccccc|}
\hline
      & $d_1$ & 0     &  1 &  2 &   3 &    4 &    5 &     6 &     7 \\
\hline
$d_2$ &       &       &    &    &     &      &      &       &       \\
0     &       &       & -2 &  0 &   0 &    0 &    0 &    0  &      0\\
1     &       &   1   & 3  &  5 &   7 &    9 &   11 &   13  &     15\\
2     &       &   0   & 0  & -6 & -32 & -110 & -288 & -644  & - 1280\\
3     &       &   0   & 0  &  0 &  27 &  286 & 1651 & 6885  &  23188\\
4     &       &   0   & 0  &  0 &   0 & -192 &-3038 &-25216 &-146718\\
5     &       &   0   & 0  &  0 &   0 &    0 & 1695 & 35870 & 392084\\
6     &       &   0   & 0  &  0 &   0 &    0 &   0  &-17064 &-454880\\
7     &       &   0   & 0  &  0 &   0 &    0 &   0  &     0 & 188454\\
\hline
\end{tabular}
\caption{Instanton numbers $n^{g=0}_{d_1d_2}$ of local $\K_{\HirzeF_1}$}
\end{table}

\begin{table}[h!]
\centering
\begin{tabular}[h]{|c|ccccccccc|}
\hline
      & $d_1$ & 0     &  1 &  2 &   3 &    4 &    5 &       6 &      7  \\
\hline
$d_2$ &       &       &    &    &     &      &      &         &         \\
0     &       &       &  0 &  0 &   0 &    0 &    0 &      0  &       0 \\
1     &       &     0 &  0 &  0 &   0 &    0 &    0 &      0  &       0 \\
2     &       &     0 &  0 &  0 &   9 &   68 &  300 &    988  &    2698 \\
3     &       &     0 &  0 &  0 & -10 & -288 &-2938 & -18470  &  -86156 \\
4     &       &     0 &  0 &  0 &   0 &  231 & 6984 &  90131  &  736788 \\
5     &       &     0 &  0 &  0 &   0 &    0 &-4452 &-152622  &-2388864 \\
6     &       &     0 &  0 &  0 &   0 &    0 &    0 &  80948  & 3164814 \\
7     &       &     0 &  0 &  0 &   0 &    0 &    0 &      0  &-1438086 \\
\hline
\end{tabular}
\caption{Genus one GV invariants $n^{g=1}_{d_1d_2}$ of local $\K_{\HirzeF_1}$}
\end{table}

\begin{table}[h!]
\centering
\begin{tabular}[h]{|c|ccccccccc|}
\hline
      & $d_1$ & 0     &  1 &  2 &   3 &    4 &    5 &       6 &       7  \\
\hline
$d_2$ &       &       &    &    &     &      &      &         &          \\
0     &       &       &  0 &  0 &   0 &    0 &    0 &      0  &         0\\
1     &       &     0 &  0 &  0 &   0 &    0 &    0 &      0  &         0\\
2     &       &     0 &  0 &  0 &   0 &  -12 & -116 &   -628  &     -2488\\
3     &       &     0 &  0 &  0 &   0 &  108 & 2353 &  23910  &    160055\\
4     &       &     0 &  0 &  0 &   0 & -102 &-7506 &-161760  &  -1921520\\
5     &       &     0 &  0 &  0 &   0 &    0 & 5430 & 329544  &   7667739\\
6     &       &     0 &  0 &  0 &   0 &    0 &    0 &-194022  & -11643066\\
7     &       &     0 &  0 &  0 &   0 &    0 &    0 &      0  &   5784837\\
\hline
\end{tabular}
\caption{Genus two GV invariants $n^{g=2}_{d_1d_2}$ of local $\K_{\HirzeF_1}$}
\end{table}

\begin{table}[h]
\centering
\begin{tabular}[h]{|c|ccccccccc|}
\hline
      & $d_1$ & 0     &  1 &  2 &   3 &    4 &    5 &       6 &       7 \\
\hline
$d_2$ &       &       &    &    &     &      &      &         &          \\
0     &       &       &  0 &  0 &   0 &    0 &    0 &       0 &         0\\
1     &       &     0 &  0 &  0 &   0 &    0 &    0 &       0 &         0\\
2     &       &     0 &  0 &  0 &   0 &    0 &   15 &     176 &      1130\\
3     &       &     0 &  0 &  0 &   0 &  -14 & -992 &  -18118 &   -182546\\
4     &       &     0 &  0 &  0 &   0 &   15 & 4519 &  179995 &   3243067\\
5     &       &     0 &  0 &  0 &   0 &    0 &-3672 & -447502 & -16230032\\
6     &       &     0 &  0 &  0 &   0 &    0 &    0 &  290853 &  28382022\\
7     &       &     0 &  0 &  0 &   0 &    0 &    0 &      0  & -15363990\\
\hline
\end{tabular}
\caption{Genus three GV invariants $n^{g=3}_{d_1d_2}$ of local $\K_{\HirzeF_1}$}
\end{table}

\clearpage


\end{document}